\documentclass[twocolumn]{aastex63}
\begin{document}

\title{2~mm GISMO Observations of the Galactic Center.
\\II. A Nonthermal Filament in the Radio Arc and Compact Sources\footnote{Based on observations carried out with the IRAM 30~m Telescope. 
IRAM is supported by INSU/CNRS (France), MPG (Germany) and IGN (Spain).}}

\shortauthors{Staguhn et al.}
\shorttitle{2~{\rm mm} Emission from the Radio Arc NTF}

\author[0000-0002-8437-0433]{Johannes Staguhn}
\affiliation{Code 665, NASA/GSFC, 8800 Greenbelt Road, Greenbelt, MD 20771, USA}
\affiliation{Johns Hopkins University}

\author[0000-0001-8403-8548]{Richard G. Arendt} 
\affiliation{Code 665, NASA/GSFC, 8800 Greenbelt Road, Greenbelt, MD 20771, USA}
\affiliation{CRESST2/UMBC}

\author[0000-0001-8033-1181]{Eli Dwek} 
\affiliation{Code 665, NASA/GSFC, 8800 Greenbelt Road, Greenbelt, MD 20771, USA}

\author[0000-0002-6753-2066]{Mark R. Morris} 
\affiliation{Department of Physics and Astronomy, University of California Los Angeles, Los Angeles, CA 90095, USA}

\author{Farhad Yusef-Zadeh} 
\affiliation{CIERA and the Department of Physics \& Astronomy, Northwestern University, 2145 Sheridan Road, Evanston, IL 60208, USA}

\author[0000-0002-9884-4206]{Dominic J. Benford} 
\affiliation{Astrophysics Division, NASA Headquarters, 300 E St. SW, Washington, DC 20546, USA}

\author[0000-0001-8991-9088]{Attila Kov\'acs} 
\affiliation{Smithsonian Astrophysical Observatory Submillimeter Array (SMA), MS-78, 60 Garden St, Cambridge, MA 02138, USA}

\author{Junellie Gonzalez-Quiles} 
\affiliation{Code 667, NASA/GSFC, 8800 Greenbelt Road, Greenbelt, MD 20771, USA}
\affiliation{CRESST2/SURA}

\email{Johannes.G.Staguhn@nasa.gov,
Richard.G.Arendt@nasa.gov}
\email{Eli.Dwek@nasa.gov,
morris@astro.ucla.edu}
\email{zadeh@northwestern.edu,
Dominic.J.Benford@nasa.gov}
\email{attila.kovacs@cfa.harvard.edu,
Junellie.Gonzalez-Quiles@nasa.gov}

\begin{abstract}
We have used the Goddard IRAM 2-Millimeter Observer (GISMO) with the 30~m IRAM telescope to carry out a 2~mm survey of the 
Galaxy's central molecular zone (CMZ). These observations detect thermal emission 
from cold ISM dust, thermal free-free emission from ionized gas, and nonthermal synchrotron 
emission from relatively flat-spectrum sources. 
Archival data sets spanning $3.6~\mu$m to 90~cm are used to distinguish different 
emission mechanisms. After the thermal emission of dust is modeled
and subtracted, the remaining 2~mm emission is dominated by free-free emission, with the 
exception of the brightest nonthermal filament (NTF) that runs though the middle of the bundle 
of filaments known as the Radio Arc. This is the shortest wavelength at which any NTF has been detected.
The GISMO observations clearly trace this NTF over a length of $\sim0.2\arcdeg$, 
with a mean 2~mm spectral index which is steeper than at longer wavelengths. The 2~mm to 
6~cm (or 20~cm) spectral index steepens from $\alpha \approx -0.2$ to $-0.7$ 
as a function distance from the Sickle \ion{H}{2} region,
suggesting that this region is directly related to the NTF.
A number of unresolved (at $21''$) 2~mm sources 
are found nearby. One appears to be thermal 
dust emission from a molecular cloud that is associated with an enigmatic radio point source
whose connection to the Radio Arc is still debated. 
The morphology and colors at shorter IR wavelengths indicate 
other 2~mm unresolved sources are likely to be compact \ion{H}{2} regions.
\end{abstract}

%\keywords{dust, extinction --- infrared: ISM --- ISM: general --- Galaxy: center
%--- radio continuum: ISM --- submillimeter: ISM}

\section{Introduction} \label{sec:intro}

\begin{figure*}[t] 
   \centering
   \includegraphics[width=7.5in]{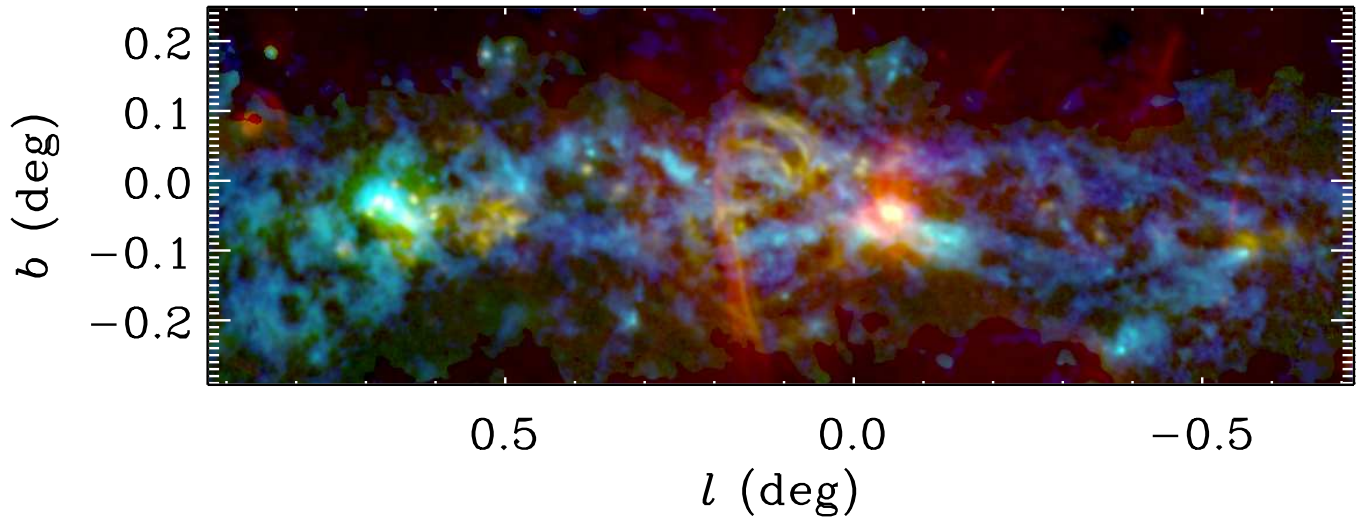}
   \caption{GISMO 2~mm observations (green) of the Galactic Center, superimposed with 
   850~$\mu$m SCUBA2 observations (blue) and 19.5~cm VLA observations (red).
   This image highlights different
   emission mechanisms from different sources. Most generally, red represents
   synchrotron emission sources, yellow corresponds to free-free emission 
   sources, and cyan features are thermal emission from dust in molecular clouds.
   High latitude regions of the 2~mm image are masked 
   where the {\it Herschel} 500 $\mu$m brightness is $<500$~MJy~sr$^{-1}$,
   affecting some areas at $b>0.1\arcdeg$ and $b<-0.2\arcdeg$. 
   All images are convolved to match the $21''$ resolution of the GISMO data.
   \label{fig:3color}}
\end{figure*}

Radio continuum observations have identified several systems of nonthermal (radio)
filaments (NTFs, sometimes known as NRFs) 
as well as isolated individual filaments within the inner two
degrees of the Galactic center, with the magnetic field aligned along the linear
filaments \citep{Yusef-Zadeh:1984, Yusef-Zadeh:1997}. A number of models have
considered the origin of NTFs associated with molecular clouds or with
mass-losing stars \citep[e.g.][]{Rosner:1996, Shore:1999, Bicknell:2001}.
However, there is no consensus on the origin of the Galactic center filaments.
Among the large number of known NTFs, a bundle of filaments near the Radio Arc
at $l\sim0.2\arcdeg$ is unique in its physical interaction with the molecular
cloud G0.13-0.13 \citep{Tsuboi:1997}. In addition, these filaments are
surrounded by ionized gas, and have an unusually flat spectrum for a 
nonthermal polarized source. Here we investigate the energy spectrum 
of the emission from the radio  arc filaments by measuring its spectral 
index at a wavelength of 2~mm, thus providing additional constraint 
on possible mixture of thermal and nonthermal emission.

The 2~mm GISMO survey of the Galactic Center provides a unique
view of thermal emission from cold dust in molecular clouds,
and free-free emission from ionized gas in \ion{H}{2} 
regions \citep[][Paper I]{Arendt:2019}. It also 
shows emission from the single brightest NTF, 
which runs through the middle of a bundle of parallel NTFs known as the Radio Arc.
While the large-scale Radio Arc at low angular resolution has been
observed to have an inverted radio spectrum ($S_\nu \propto \nu^\alpha$ 
with $\alpha > 0$) at frequencies as high
as 43~GHz (7~mm) \citep{Reich:1988}, the smaller-scale structure has been reported to have
a steep ($\alpha \approx -1$) spectrum between 32 and 43~GHz (or 7 to 9~mm) \citep{Sofue:1999a}.
The bright NTF had been detected in interferometric observations at wavelengths as short as 3~mm 
\cite{Pound:2018} and polarized emission has been detected from the entire Radio
Arc at low resolution at 2 and 3~mm \citep{Culverhouse:2011}.
Most NTFs across the GC have relatively steep radio spectra \citep[e.g.][]{Law:2008a},
and any emission at 2~mm or shorter wavelengths would be expected to be very 
faint and confused. The NTFs had not been detected at shorter wavelengths (e.g. 850~$\mu$m 
and 1.1~mm), which made emission from these features somewhat unexpected 
at 2~mm as previously reported by \cite{Reich:2000}.

The 2~mm wavelength is a transition region in the spectrum of the general Galactic ISM 
and specific Galactic sources. The Rayleigh-Jeans tail of thermal emission from cold dust 
continues to fade from far-IR wavelengths into the 2~mm band, whereas 
optically thin free-free (thermal) radio sources are fading as wavelengths fall from the 
radio regime to the mm and sub-mm regime. Optically thin non-thermal 
sources typically fall much more quickly as a function of decreasing wavelength.

Figure \ref{fig:3color} provides an overview of the 2~mm emission (green) in comparison
with the 850~$\mu$m thermal emission from dust \citep[blue;][]{Parsons:2018} 
and 19.5~cm radio emission (red).
Most molecular clouds are identified by cyan colors with dust emission extending
from 850~$\mu$m to 2~mm. Regions of radio emission with relatively flat spectral 
indices (free-free emission and flat non-thermal sources) exhibit yellow colors
in this image. 

In this paper, we employ a wide wavelength range of archival data sets 
to focus on the analysis of the emission from the brightest of the NTFs. 
We investigate if the 2~mm observation can provide further insight 
into the origin and nature of the NTFs, and their relation to other possibly nearby 
features. In particular, there is a radio point source of an uncertain nature that
lies atop the NTF, at least in projection. We find 2~mm emission 
at the location of this source and compare the SED of this 2~mm point source with 
other nearby point-like 2~mm sources in order to understand the nature of this source.

\section{Data} \label{sec:data}

The GISMO instrument \citep{Staguhn:2006,Staguhn:2008}, 
paired with the 30~m IRAM telescope \citep{Baars:1987},
was used to map a $\sim 2\arcdeg \times 0.6\arcdeg$ region 
at the Galactic Center at a wavelength of 2~mm. 
The beam size of GISMO observations at the 30~m telescope is $16.6''$ FWHM. 
The observations were generally carried out under stable atmospheric 
conditions and with a zenith opacity of $\tau_{\rm 2mm} < 0.11$. Focus in 
z-direction was regularly monitored (several times a day) and pointing 
was frequently checked (about once per observing hour) using the nearby 
bright quasars J2037+511 and J1637+574. Fluxes were calibrated to $<10\%$ 
accuracy by monitoring Mars and employing the atmospheric transmission 
model of the Caltech Submillimeter 
Observatory\footnote{\url{http://www.submm.caltech.edu/cso/weather/atplot.shtml}}
and the 30~m telescope 225 GHz radiometer readings.
The data reduction and 
mosaicking, done with CRUSH \citep{Kovacs:2008}, yields an image with $21''$ (FWHM) resolution.

\citet{Arendt:2019} describe the characterization of the thermal 
dust emission in the image via the extrapolation of a modified blackbody spectrum
based on 160 - 500~$\mu$m {\it Herschel} PACS and SPIRE observations \citep{Molinari:2016}. 
The Herschel data, convolved to $37''$ resolution, were initially 
fit on a pixel-to-pixel basis to determine a mean dust temperature, 
a spectral index, $\beta$, of the dust emissivity, $\kappa_\nu(\lambda)$, and a normalization 
(proportional to the mass). These fits occasionally produced spurious 
results, so a constrained model was applied in which the dust emissivity was 
characterized as $\kappa_\nu(\lambda) = \kappa_0 \lambda^{-2.25}$, where the 
exponent $\beta = -2.25$ was determined from the mean of the values
found in the unconstrained fit. [See \citep{Arendt:2019} for details.]
This modeling is used to subtract the thermal dust component from the 2~mm 
emission, leaving the free-free emission from regions of ionized gas, and
non-thermal (synchrotron) emission from sources that combine relativistic electrons
with strong magnetic fields. 

Additional Galactic center maps used here are: 850~$\mu$m SCUBA-2 observations
from \cite{Parsons:2018}; 1.1~mm observations from the 
BOLOCAM Galactic Plane Survey \citep{Ginsburg:2013};
{\it Herschel} PACS and SPIRE observations from the Hi-GAL survey \citep{Molinari:2016};
and {\it Spitzer} IRAC and MIPS observations at 3.6 - 24~$\mu$m 
\citep{Stolovy:2006,Yusef-Zadeh:2009}. 
VLA observations at 6 and 20~cm are from 
\cite{Yusef-Zadeh:1984}, \cite{Yusef-Zadeh:1987,Yusef-Zadeh:1987a},
and \cite{Yusef-Zadeh:2009}.
VLA observations at 90~cm are from \cite{LaRosa:2000}.

\begin{figure*}[p] 
   \centering
   \includegraphics[width=3.5in]{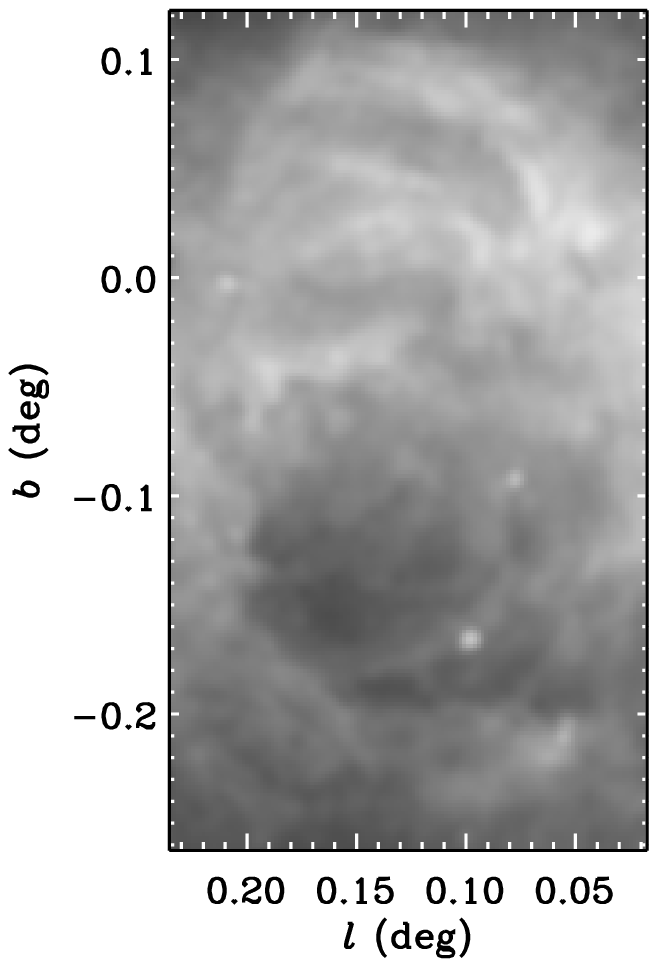}
   \includegraphics[width=3.5in]{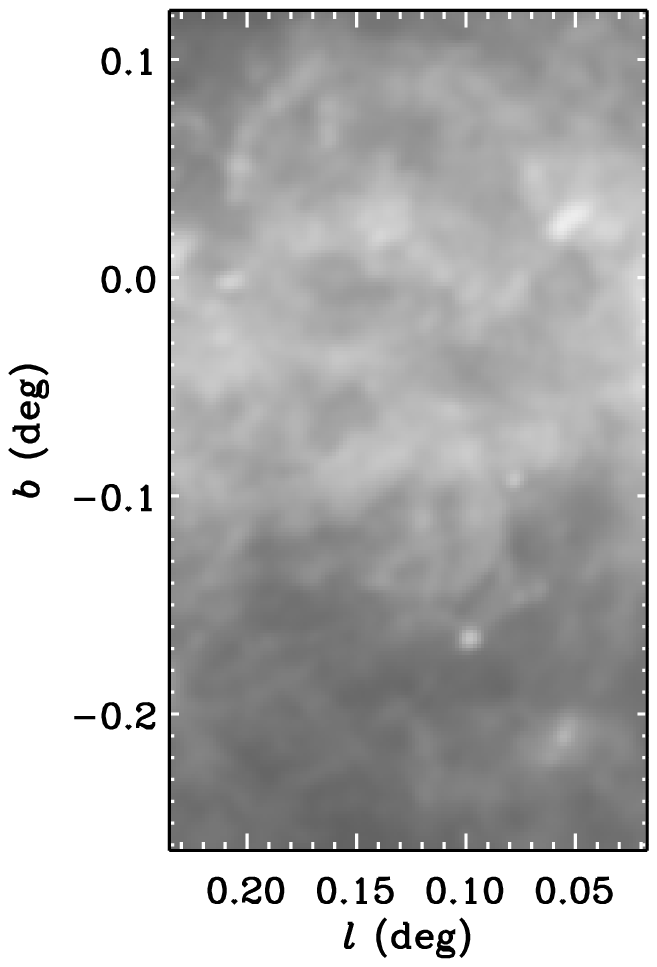}\\
   \includegraphics[width=3.5in]{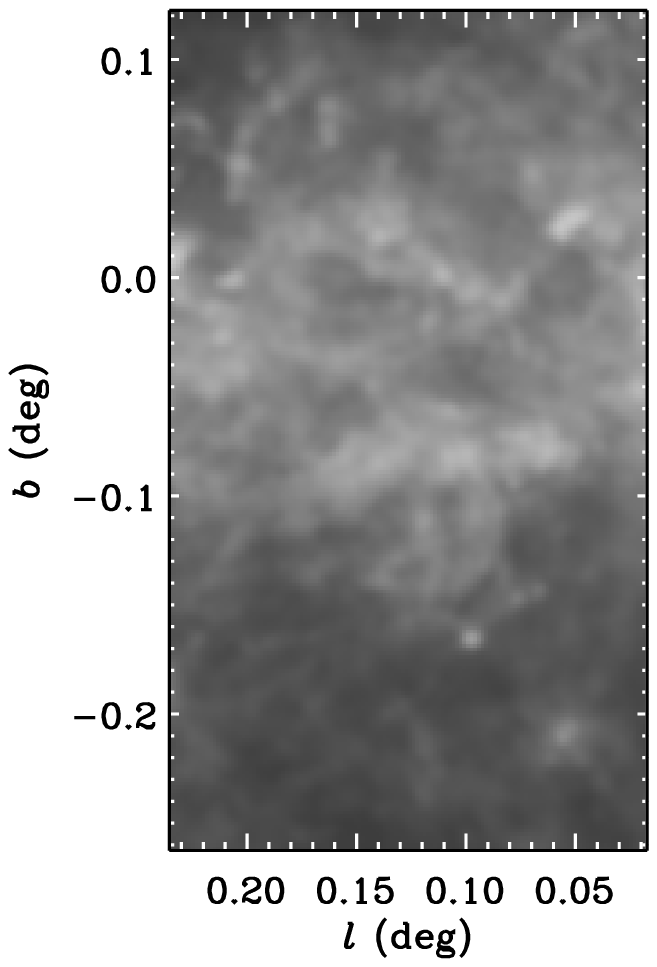}
   \includegraphics[width=3.5in]{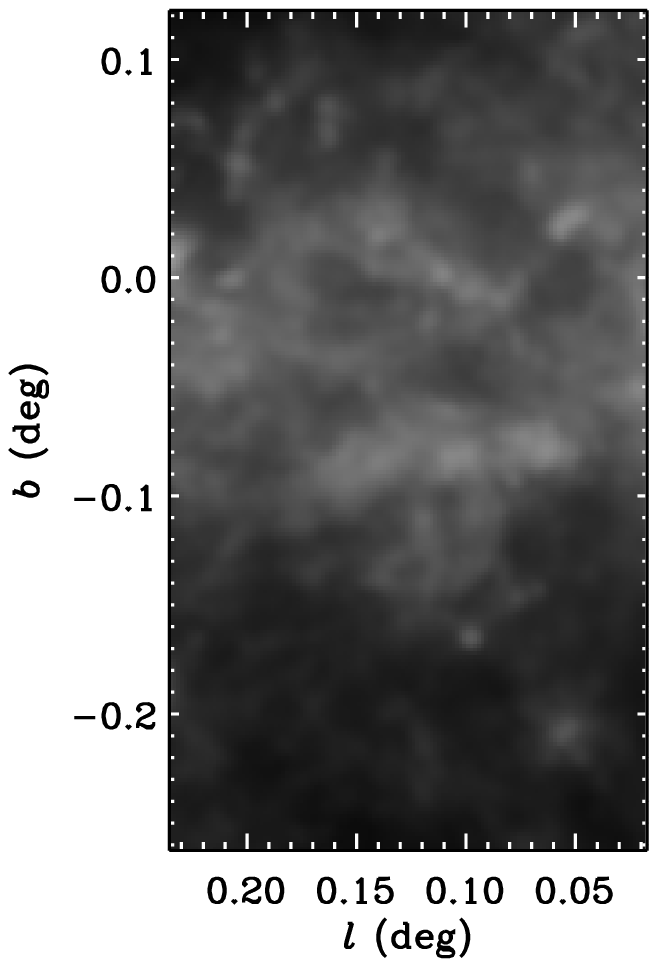}
   \caption{{\it Herschel} PACS (70, 160~$\mu$m, left to right, top row) 
   and SPIRE (250, 350 $\mu$m, left to right, bottom row) images of the 
   Radio Arc region. The data are displayed on a logarithmic scale 
   ($10^3$ -- $10^5$~MJy sr$^{-1}$) and convolved to $21''$ resolution to match the GISMO data.
   \label{fig:pacs}}
\end{figure*}

\begin{figure*}[p]
   \centering
   \includegraphics[width=3.5in]{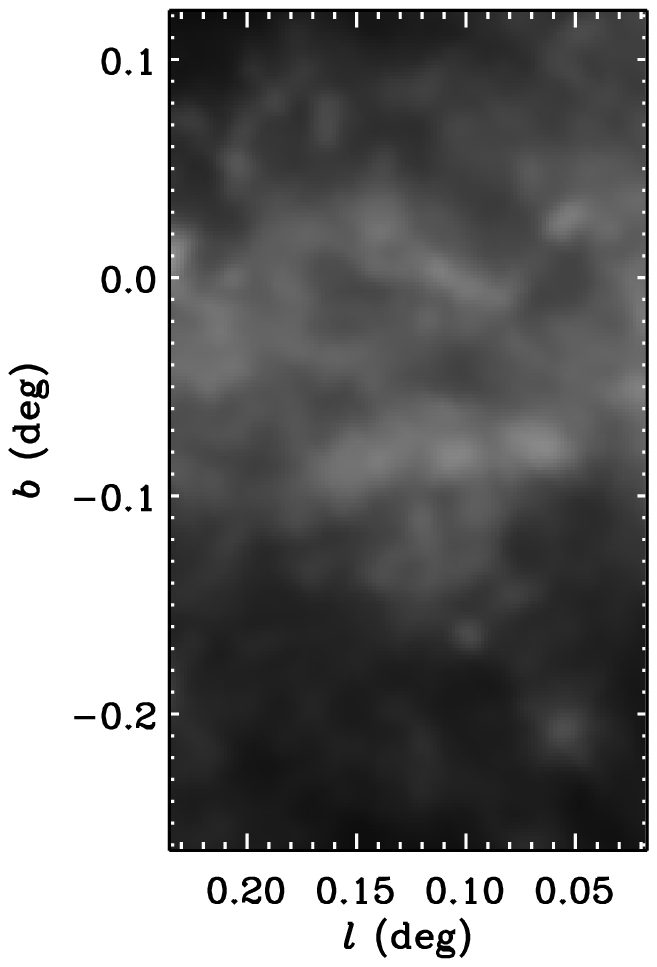} 
   \includegraphics[width=3.5in]{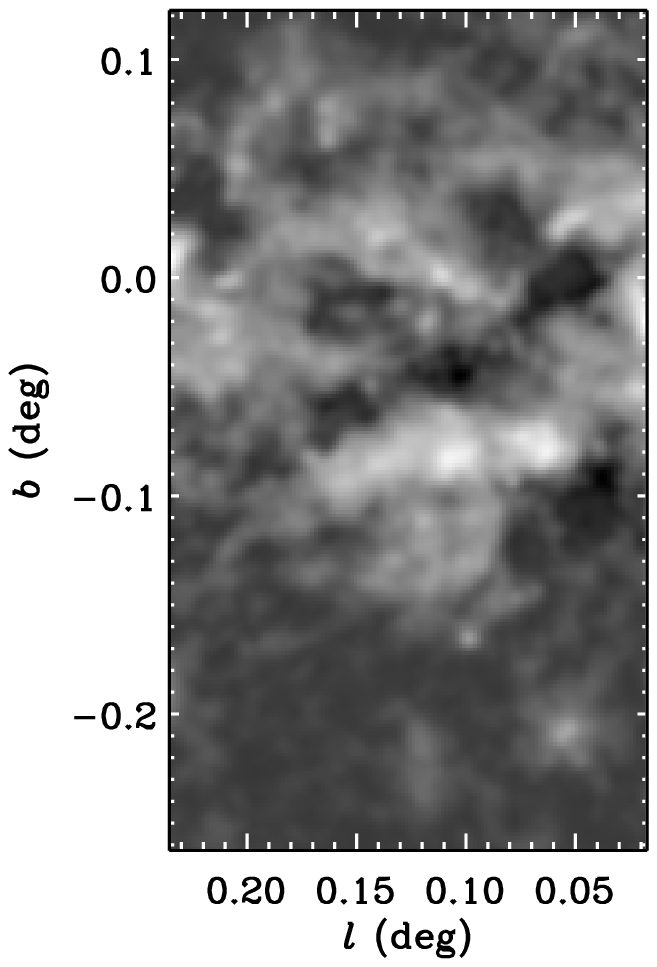}\\
   \includegraphics[width=3.5in]{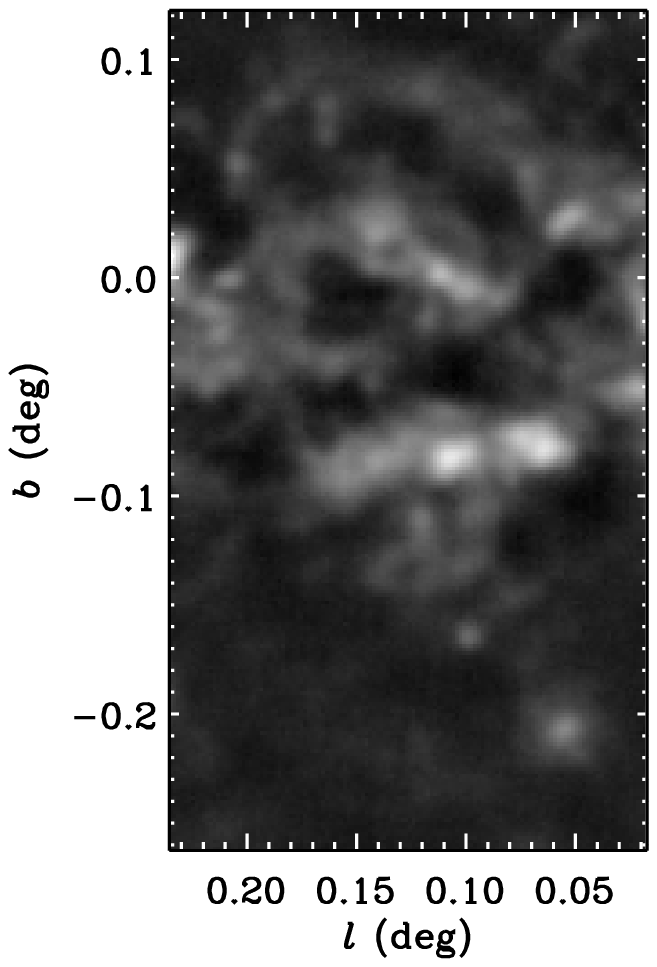}
   \includegraphics[width=3.5in]{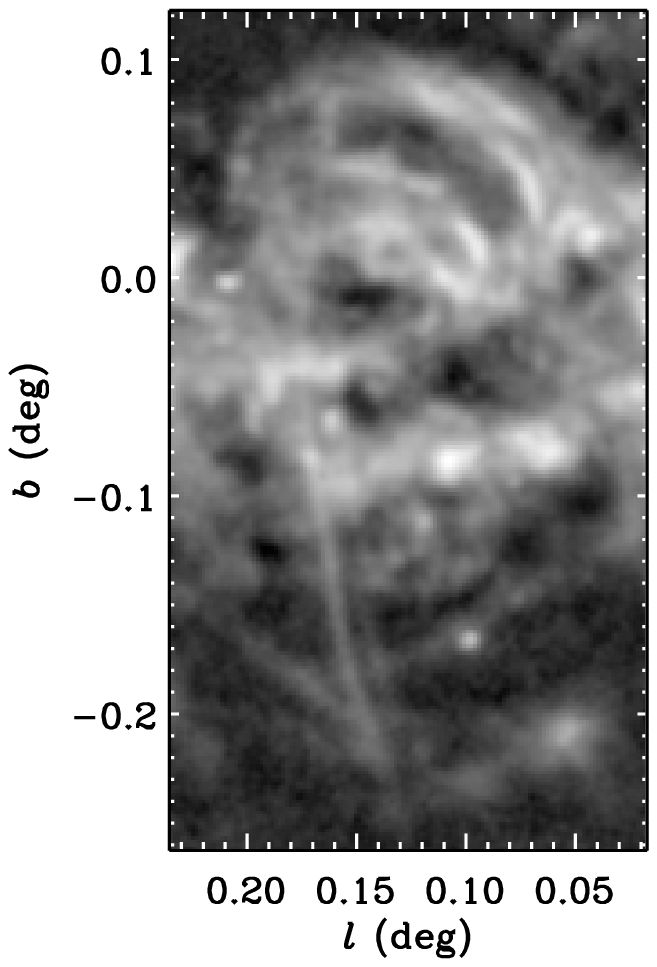} 
   \caption{{\it Herschel} SPIRE (500 $\mu$m, top left) image of the Radio Arc region. SCUBA-2 (850~$\mu$m, top right), 
   BOLOCAM (1.1~mm, bottom left), and GISMO (2~mm, bottom right) images. All images are logarithmically scaled: 
   300 -- $3\times 10^4$~MJy sr$^{-1}$ for 500~$\mu$m, minimum to maximum for the other images after adding an offset 
   approximately equal to the minimum value.  Image resolutions are, respectively, $37''$, $21''$, $30''$, and $21''$.
   \label{fig:spire}}
\end{figure*}

\begin{figure*}[p] 
   \centering
   \includegraphics[width=3.5in]{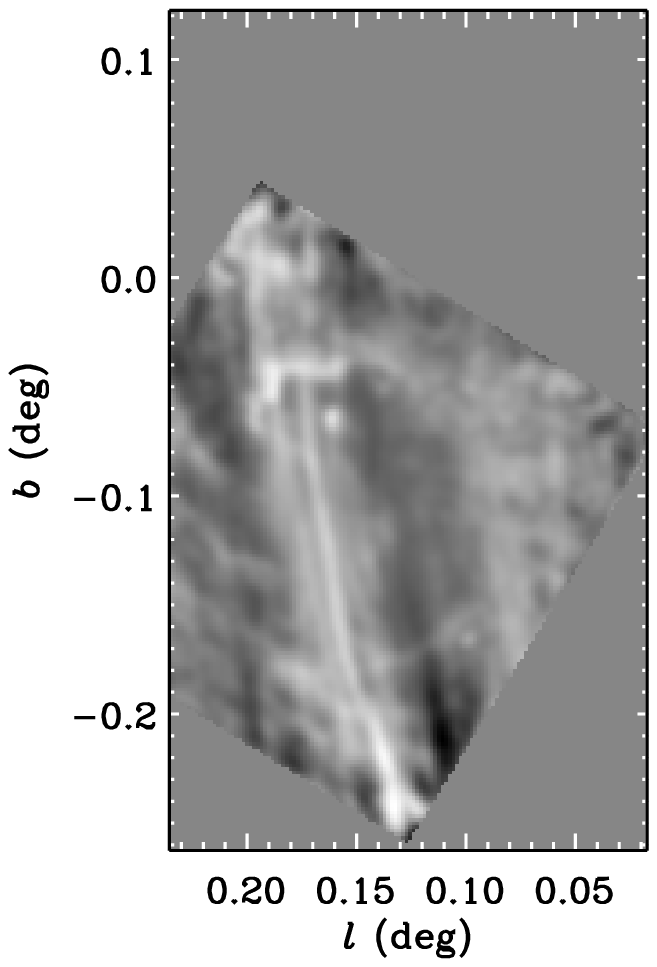}
    \includegraphics[width=3.5in]{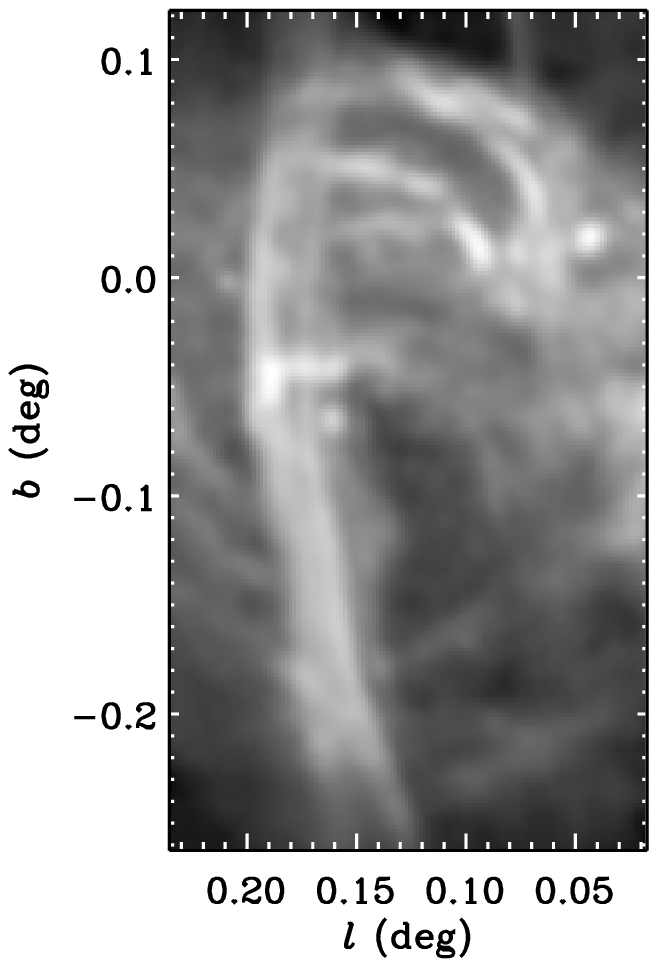}\\
   \includegraphics[width=3.5in]{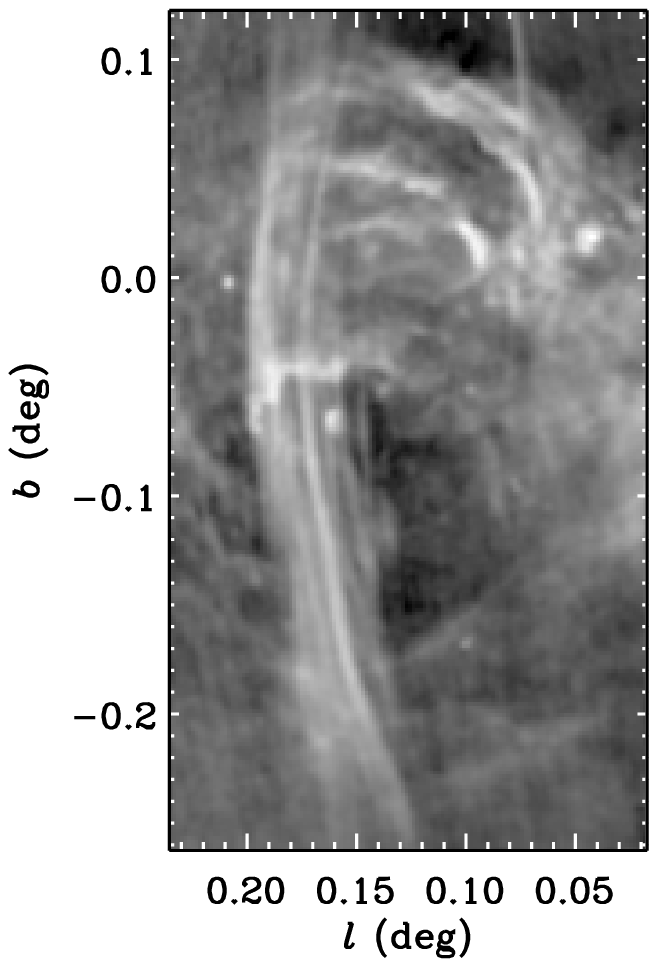}
   \includegraphics[width=3.5in]{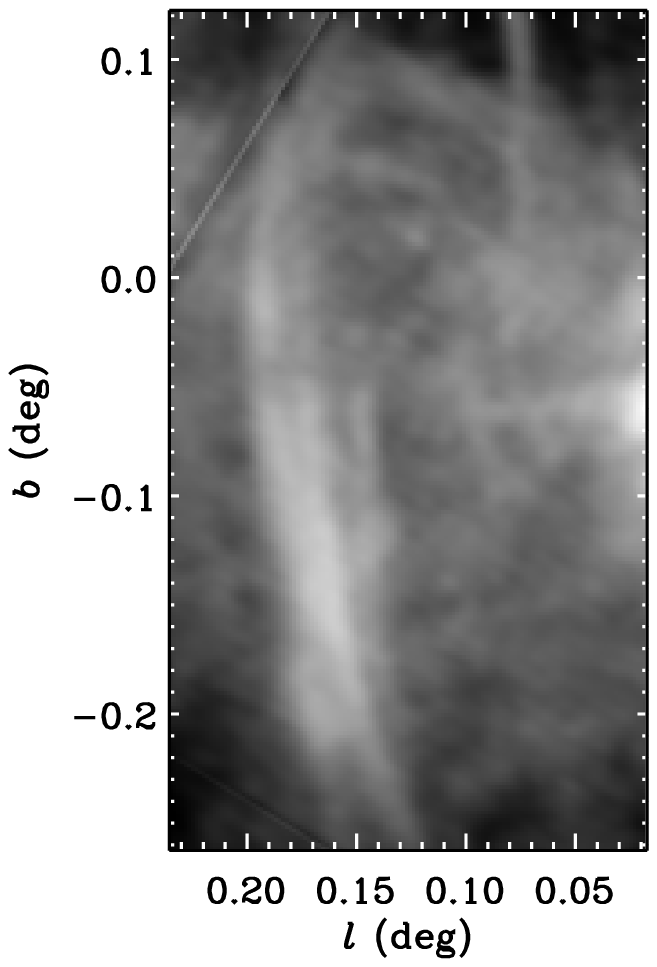} 
   \caption{VLA 6.2, 19.5, 20.1, and 90 cm images of the Radio Arc region (left to right, top to bottom). 
   All images are logarithmically scaled, minimum to maximum after adding an offset 
   approximately equal to the minimum value. 
   Image resolutions are $21''$, $30''$, $21''$, and $43.2'' \times 23.6''$ respectively.
   The diagonal line in the upper left corner of the 90~cm image is a mosaicking artifact.
   \label{fig:radio}}
\end{figure*}

Figures \ref{fig:pacs} and \ref{fig:spire} illustrate the emission from the region 
of the Radio Arc at wavelengths from 70~$\mu$m to 2~mm. As wavelength increases,
emission from warm dust becomes surpassed by emission from colder dust, 
which in turn begins to be exceeded by free-free emission at 2~mm.
Figure \ref{fig:radio} further illustrates emission at radio wavelengths from
6 to 90~cm. The primary beam correction \citep{Napier:1982} applied to the 6~cm 
image appears valid over the range $-0.2\arcdeg<b<0\arcdeg$, but may overcorrect
in the corners of the image. 
The 6~cm image is largely dominated by free-free emission. This gradually 
becomes dominated by non-thermal emission as wavelengths increase to 90~cm.
At 90~cm some of the free-free emission becomes optically thick,
such as at the Sickle, $(l,b)=(0.18,-0.04)$, which is seen in absorption 
against the Radio Arc filaments 
\citep{Anantharamaiah:1991}. However since our focus is on much 
shorter wavelengths, we assume that free-free emission is optically 
thin throughout the rest of this paper.

Figure \ref{fig:res} compares the 2~mm emission before and after the modeling
and subtraction of the extrapolated dust emission. In the dust-subtracted 
image, the remaining features are much more clearly associated with 
radio emission features at $\lambda \geq 6$~cm. The third panel in the figure
highlights the extended and point-like sources that are discussed in the 
following sections.

\setcounter{figure}{4}
\begin{figure*}[p] 
   \centering
   \includegraphics[width=3.5in]{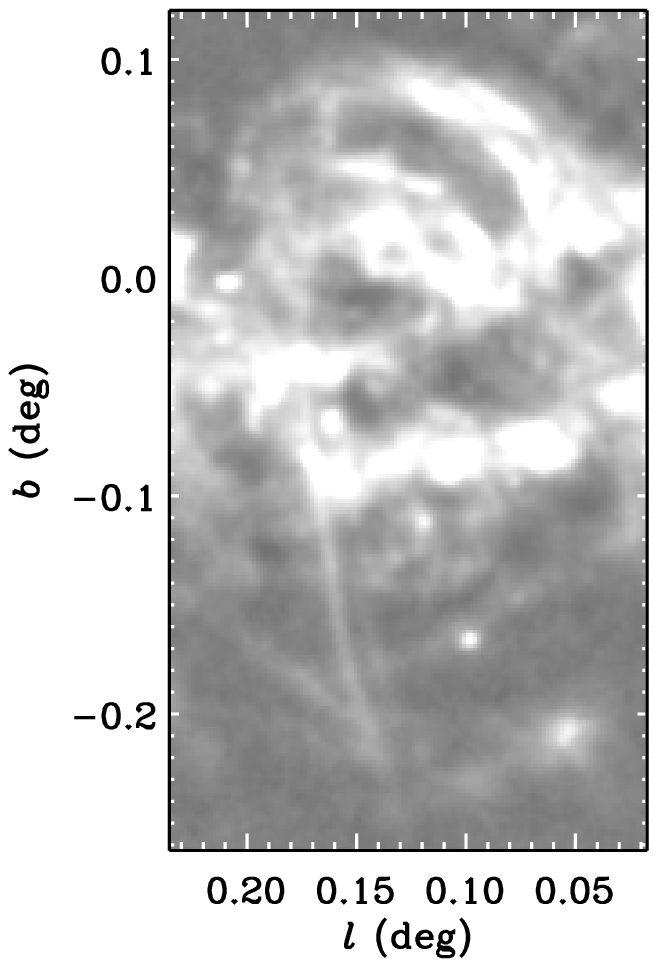} 
   \includegraphics[width=3.5in]{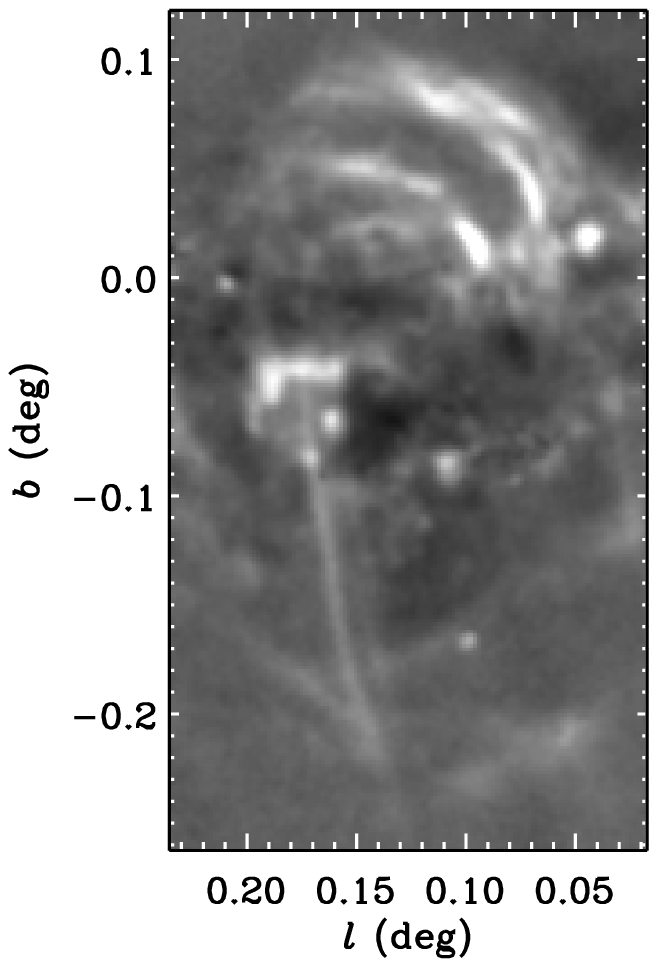}\\
   \includegraphics[width=3.5in]{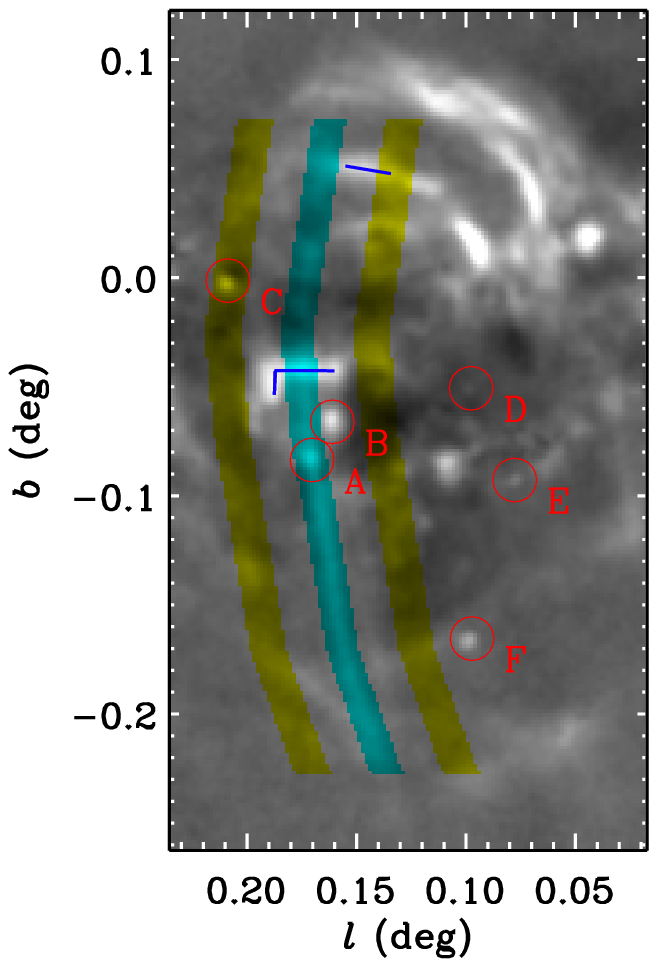} 
   \includegraphics[width=3.5in]{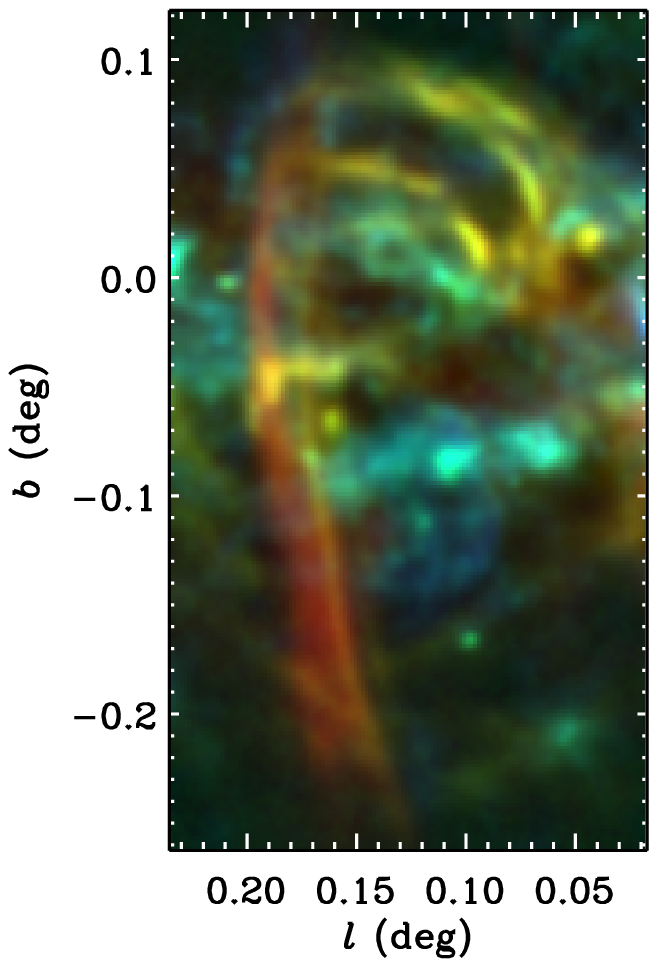}\\
   \caption{Images of the Radio Arc region at 2~mm before (top left) and 
   after (top right) subtraction of 
   thermal emission from cold dust, extrapolated from {\it Herschel} observations. 
   Both images are scaled linearly from $-10$ to 10~MJy sr$^{-1}$.
   The long cyan band in the lower left panel indicates the region integrated to 
   calculate the brightness profile of the NTF, and the yellow bands show the regions 
   used to define the background level.
   The shorter blue lines indicate regions where we examined the emission of 
   the Sickle [the L-shaped feature at $(l,b) = (0.18, -0.04)$] 
   and a portion of the Arches (the complex of bright arcs at $0.05<l< 0.20$, $0.00<b<0.12$) 
   for comparison. The red circles indicate the 
   locations of 6 compact 2~mm and radio sources that we examined. 
   The lower right panel shows 850~$\mu$m (blue), 2~mm (green), and 19.5~cm (red), 
   all linearly scaled. Different regions with different emission 
   mechanisms are distinguished by distinct colors here.
   \label{fig:res}}
\end{figure*}

%===================================
\section{Analysis} \label{sec:analysis}
%===================================

\subsection{The non-thermal filament}

\begin{figure}[t]
   \centering
   \includegraphics[width=3.5in]{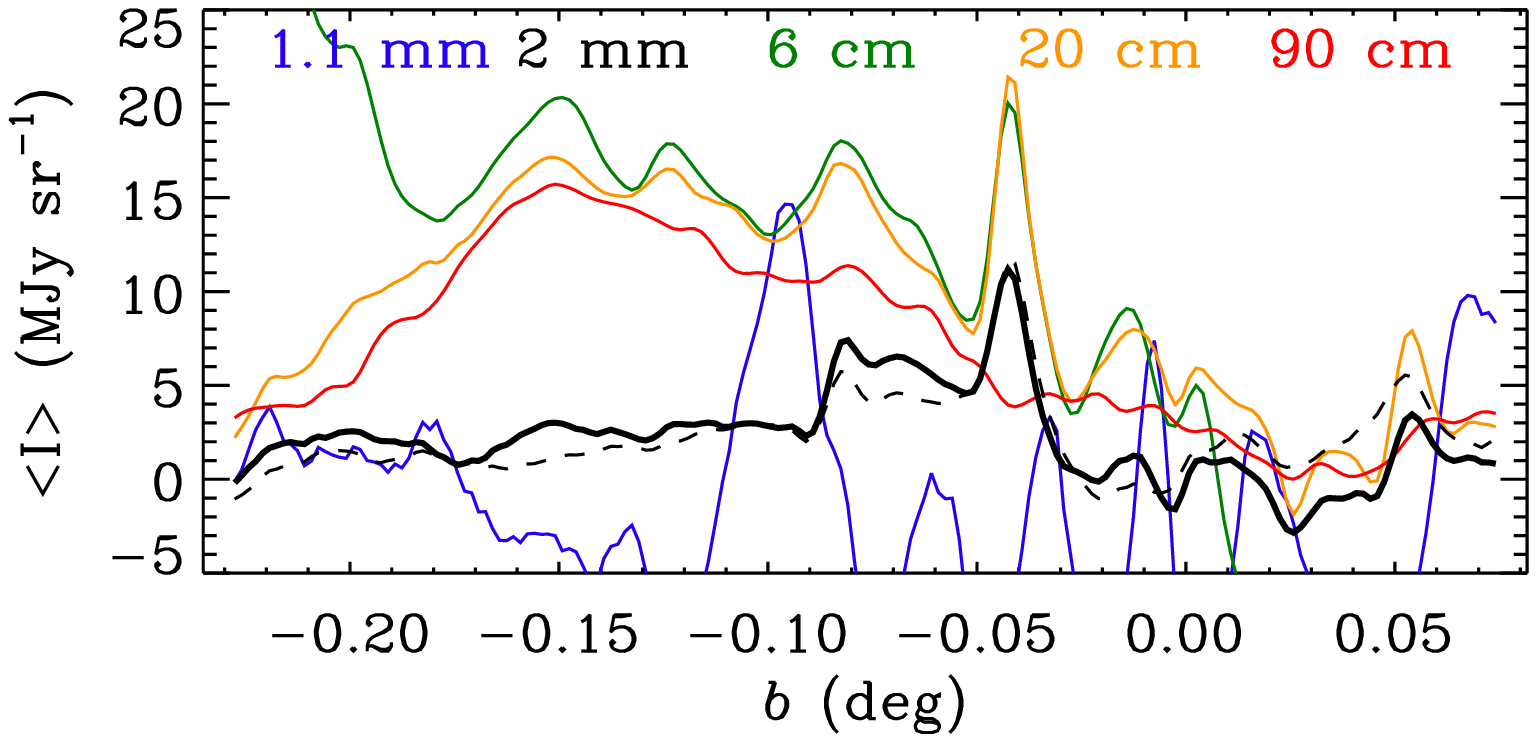} \\
   \includegraphics[width=3.5in]{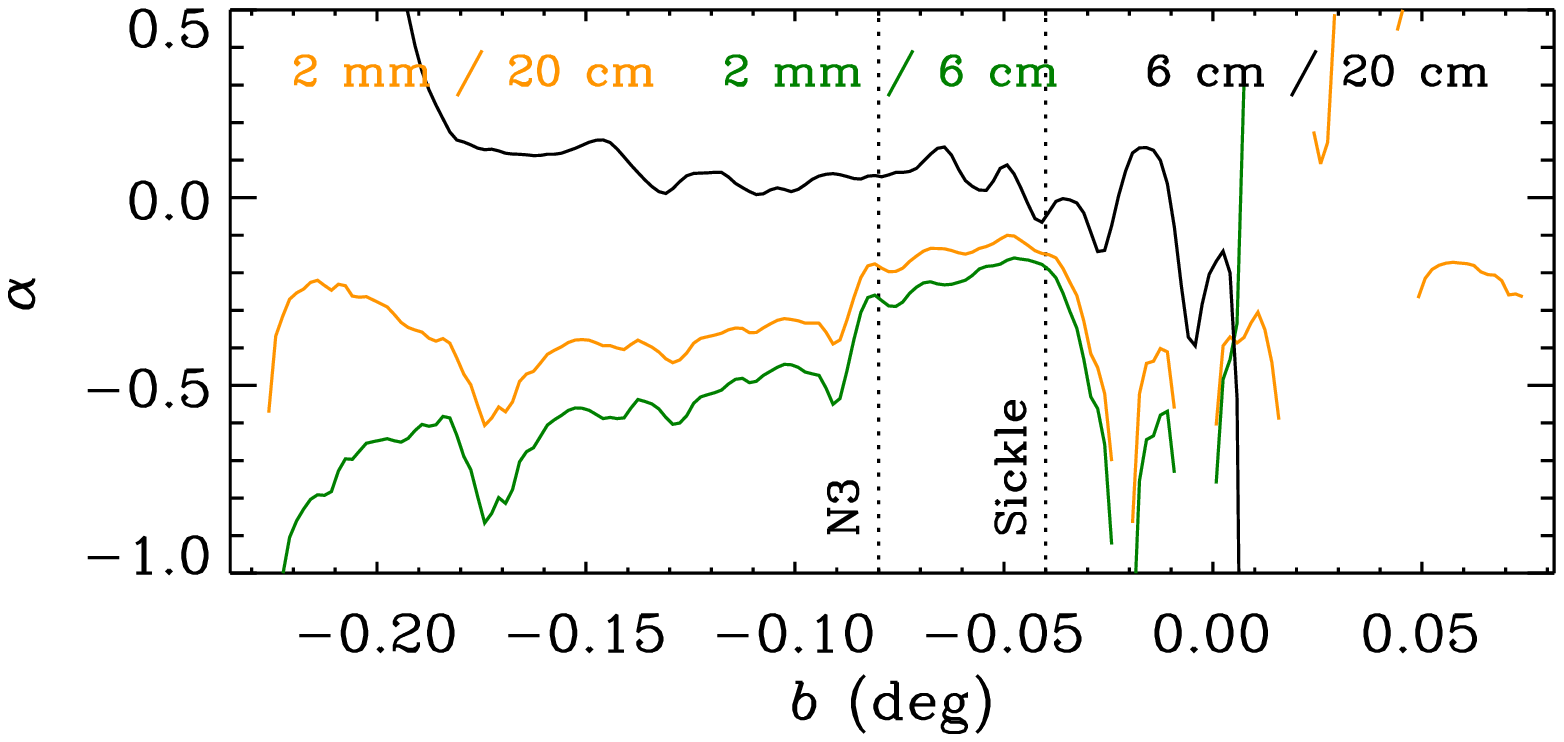}
   \caption{Mean intensity (top) and spectral index (bottom) along the main filament.
      The dashed black line in the intensity plot shows the 2~mm emission in the case 
      where the background is taken to be constant along the length of the filament 
      rather than varying as a function of latitude. The locations where the filament crosses 
      the Sickle and the N3 source (and its related molecular cloud) are indicated in the
      spectral index plot.
   \label{fig:profile}}
\end{figure}

The first step in the analysis of the emission from the NTF is to 
extract the brightness profile along the filament. A 9-pixel ($54''$) 
wide region is averaged at each latitude. A local background is subtracted 
by averaging 10-pixel ($60''$) wide strips located 
15 pixels ($90''$) east and west of the center of the NTF. 
The backgrounds are taken at the same fixed latitudes as each 
point along the NTF. The separation places the background regions 
outside the main bundle of filaments in the Radio Arc. 
The mean intensity profiles at 2~mm (after dust emission removal), 
radio wavelengths, and 1.1~mm 
are shown in Figure \ref{fig:profile}. 
The main uncertainty in the measurement of the filament brightness is
the subtraction of the background. 
The dashed line shows the 2~mm brightness profile using a 
background that is flat, averaged over all latitudes, and provides an 
indication of the amount of variation that
may occur due to confusion. This comparison demonstrates that 
some of the apparent brightness changes in the filament may be 
caused by a mismatch in the brightness of the chosen background regions 
and the actual brightness underlying the corresponding location of 
the filament. The filament is not detected at 1.1~mm. The brightness 
profile at that wavelength shows variations that may be $\sim5$ times
larger than the brightness of the filament, if it is comparable to that at 2~mm.

The main region where the 2~mm emission of the NTF 
is bright and unconfused spans from 
$-0.16\arcdeg < b <-0.09\arcdeg$,
which is roughly from the southern edge of the Radio Arc Bubble 
\citep[seen as a $\sim0.2\arcdeg$ diameter ring centered at $(l,b) = (0.15\arcdeg, -0.1\arcdeg)$][]
{Levine:1999,Rodriguez-Fernandez:2001,Simpson:2007} to the N3 radio point source 
at $(l,b) = (0.17\arcdeg, -0.08\arcdeg)$ \citep{Yusef-Zadeh:1987,Ludovici:2016} which 
is also coincident with the 2~mm source ``A'' indicated in Figure \ref{fig:res}.
To the north of this, between N3 and the Sickle ($b = -0.04\arcdeg$), the background becomes
brighter and more complex, making it difficult to determine the truly appropriate
background for the NTF. To the north of the Sickle, the NTF seems to fade
and become confused with the bright thermal emission of the Arches. The apparent brightness
of the filament in this region also seems to be diminished by the extrapolation and 
subtraction of the thermal emission from dust, even though the NTF is not 
identifiable in the far-IR emission.

The resulting profiles of the spectral indices ($\alpha$, defined by $I_{\nu} \propto \nu^{\alpha}$) 
are shown in Figure \ref{fig:profile}. In the 
relatively clean region, $-0.16\arcdeg < b < -0.9\arcdeg$, the 2~mm spectral indices seem to show a 
slight steepening (become more negative) as a function of distance south from the Sickle 
(at $b = -0.04\arcdeg$). The dip in the 2~mm spectral indices at $b = -0.09\arcdeg$ may be an
artifact of imperfect subtraction of dust emission from the bright molecular cloud
that overlaps the NTF here (see Figure \ref{fig:res}).
The 6-20~cm spectral index lies in the range $0<\alpha<0.2$, which 
is consistent with the results found by \cite{Wang:2002} in 
slices across the Radio Arc near N3. It also is similar to the spectral 
index measured in the northernmost $8'$ ($b>0.15\arcdeg$)  
portion of the Radio Arc by \cite{Law:2008a}, who note a steepening of the 
6-20~cm spectral index with increasing distance from the Galactic plane.

\begin{figure}[t] 
   \centering
   \includegraphics[width=3.5in]{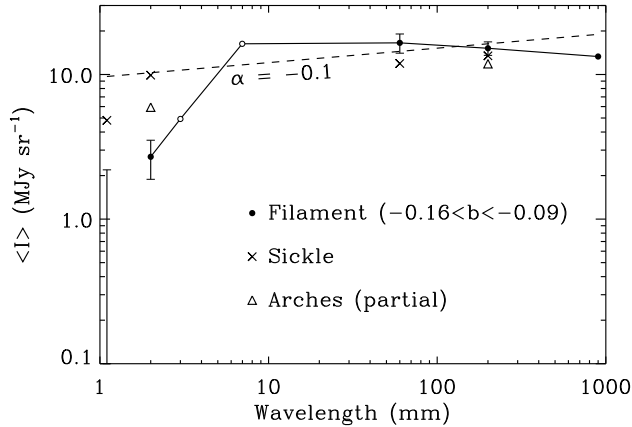}
   \caption{Mean intensity of the main filament for $-0.16\arcdeg < b < -0.09\arcdeg$.
   The 1.1~mm BOLOCAM data provide only an upper limit (1$\sigma$) on the mean intensity.
   The 2~mm brightness is measured after subtraction of the extrapolated thermal emission from dust.
   The 3 and 7~mm data points (without uncertainties) are from \cite{Pound:2018} and \cite{Sofue:1992}.
   Spectra of the Sickle and a portion of the Arches (nearest the filament) are shown 
   as examples of free-free emission sources. An arbitrarily scaled free-free spectrum 
   ($I \propto \nu^{-0.1}$) is shown for reference.
   \label{fig:filament_sed}}
\end{figure}

Figure \ref{fig:filament_sed} shows the SED averaged over the 
$-0.16\arcdeg < b < -0.09\arcdeg$ range. The 3~mm point is the interferometric measurement
of \cite{Pound:2018}, which may be considered a lower limit. 
\cite{Sofue:1992} do not reliably detect the NTF in 7~mm interferometric observations, 
but they quantify the clear detection in lower-resolution single-dish 7~mm 
observations \citep{Sofue:1986,Reich:1988} at the level shown in the figure.
There is a clear steepening of the spectrum at 2 and 3~mm relative to that at $\lambda > 6$~cm,
such that the spectral index is $\alpha \lesssim -1.5$ at 2 - 3~mm.
For comparison, we similarly extracted SEDs for known thermal sources: the Sickle and 
a portion of the Arches. In these cases, a surface brightness limit was used to 
define an aperture for each source, and a background was subtracted using the 
same aperture displaced to relatively faint regions $108''$ north of the Sickle 
and $60''$ north of the Arches segment, respectively. 
The 1.1~mm result for the NTF should be interpreted as an upper limit
on the emission rather than a detection.
The 6~cm image used here does not cover the Arches. 
At 90 cm, the Sickle is in absorption, and the Arches are too confused for a 
reliable brightness measurement in the chosen portion.

\subsection{Compact sources}

After subtraction of the extrapolated dust emission, a number of compact 2~mm sources
remain in the region of the Radio Arc. Most notably, there is a source (``A'' in Figure \ref{fig:res})
which appears to lie on the NTF at the location of the radio source N3. The radio source 
N3 is very enigmatic, and its physical association with the Radio Arc is unclear.
\cite{Ludovici:2016} provide a detailed look at the source, ruling out several possible explanations,
leaving a micro-blazar as the remaining best guess.

\begin{figure}[t] 
   \centering
   \includegraphics[width=3.2in]{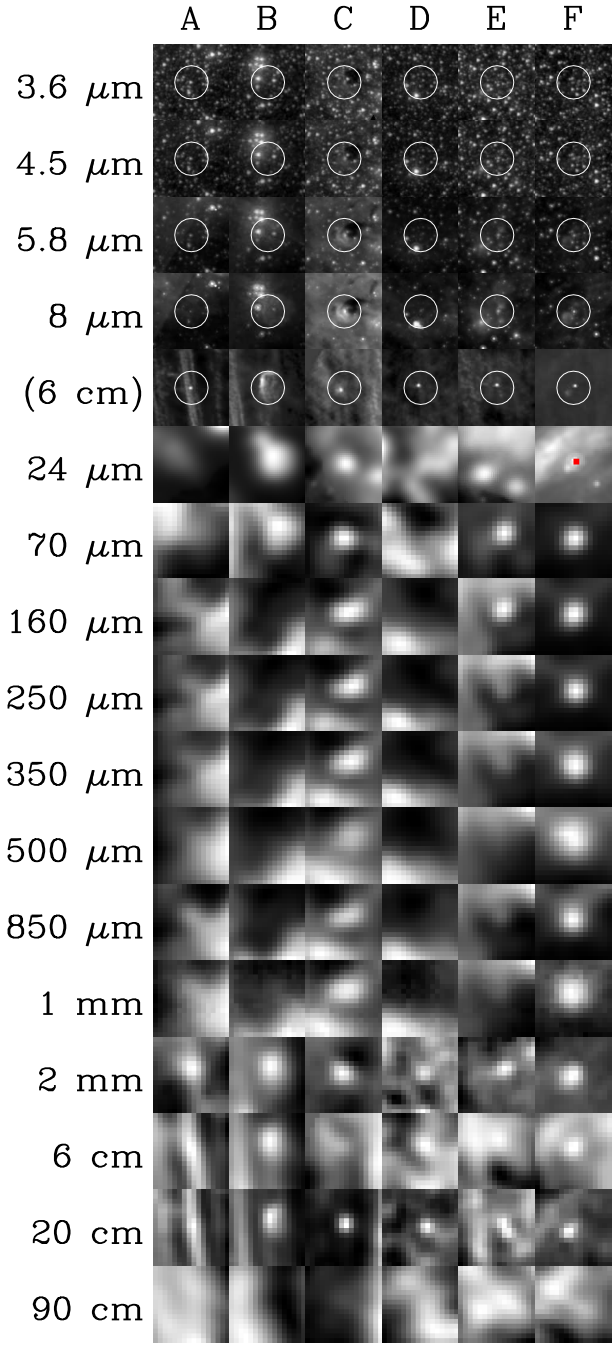} 
   \caption{Cut-out images ($96''\times96''$) of 2~mm point sources 
   (from Figure \ref{fig:res} and Table \ref{tab:sources}) at various wavelengths. 
   Source A is on the brightest NTF. Source B is the Pistol Nebula.
   Saturated pixels in the MIPS 24 $\mu$m image of source F are marked in red.
   The 24 $\mu$m MIPS images of the other sources are more severely saturated, 
   and thus we show lower resolution {\it MSX} data as presented by 
   \cite{Yusef-Zadeh:2009}. The $48''$ diameter circles in the higher resolution
   images are guides to the location of potential counterparts in
   these images.
   \label{fig:points}}
\end{figure}

\begin{figure*}[t] 
   \centering
   \includegraphics[width=7.25in]{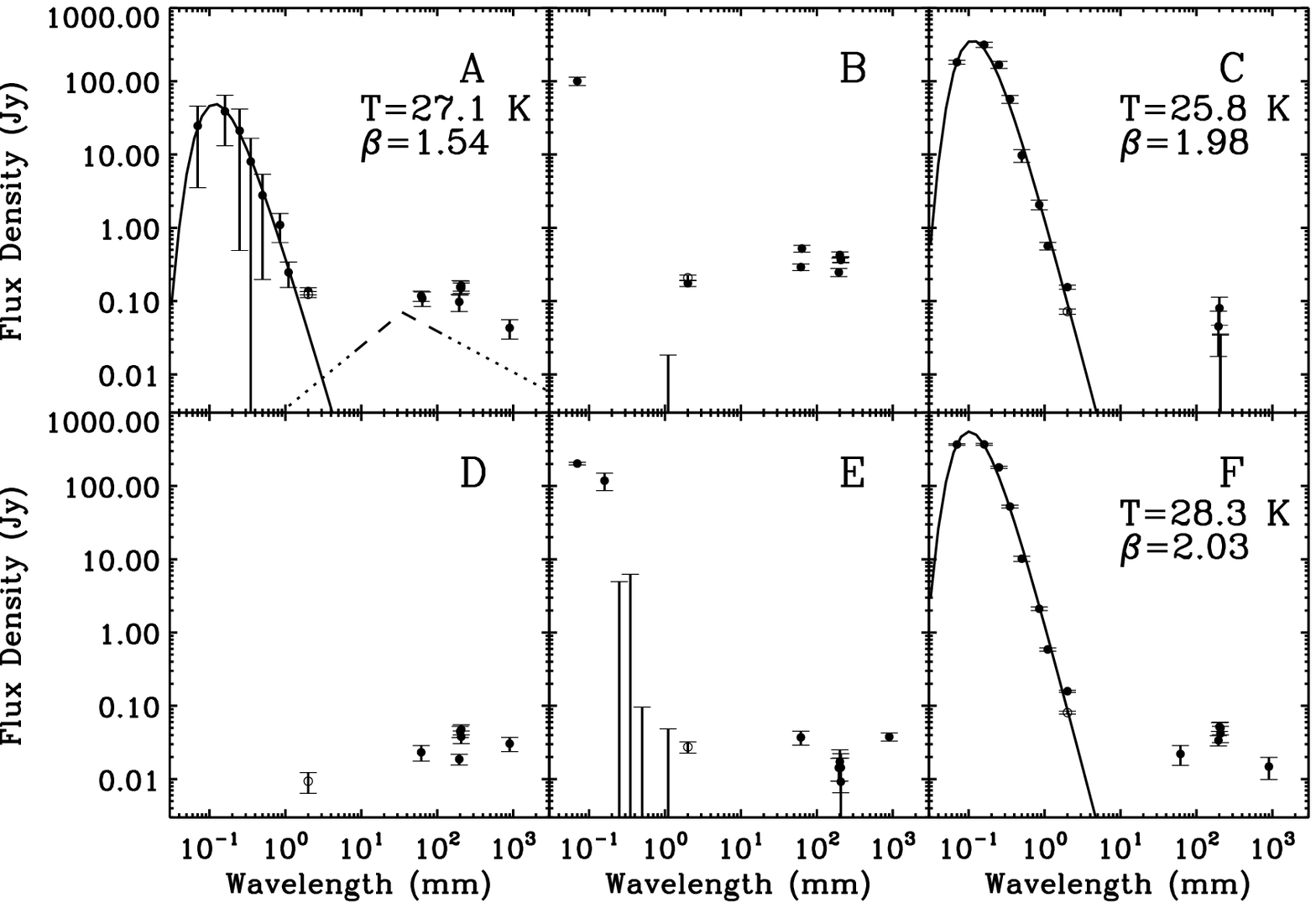} 
   \caption{Aperture photometry SEDs of 2~mm point sources (from Fig. \ref{fig:res}). 
   Pairs of 2~mm data points are measurements before (filled circle) and after (open circle) subtraction of the 
   estimated dust emission which can affect both source and background.
   Modified blackbody spectra were fit to the far-IR emission of 
   sources A, C, and F. The dashed line in panel A shows the fit to the radio SED of Source N3, using 
   other high-resolution  data sets, as determined by \cite{Ludovici:2016}. 
   The dotted lines are extrapolations of this fit. 
   The measured flux densities are available as a machine readable table.
   \label{fig:point_spectra1}}
\end{figure*}

Here, we take a further look at the spectral properties of source A, in comparison to 
the other nearby compact 2~mm sources. 
The other nearby sources are simply the nearest 2~mm point-like sources, chosen
as a comparison sample of sources that are likely to be typical compact sources.
Strong similarity of source A to nearby sources would add weight to the argument that 
it is only coincidentally along the line of sight to the NTF.
The sources indicated in Figure \ref{fig:res} and listed in Table~\ref{tab:sources} all 
appear compact ($\lesssim 21''$ FWHM) in the residual 2~mm image, and all appear as 
longer-wavelength radio sources as well. Figure \ref{fig:points} shows cut-out images of each of the 
sources as observed in the various 3.6~$\mu$m -- 90~cm data sets described in Section 2.
Figure \ref{fig:point_spectra1} displays the SEDs at far-IR to radio wavelengths using aperture
photometry. 
A circular $48''$ diameter aperture is used for the source, and a $48''-96''$ diameter
annulus is used for the background at each wavelength. 
Uncertainties are calculated from standard propagation of errors assuming that the 
uncertainty for each pixel is given by the measured standard deviation of pixel intensities in 
the background annulus.

\begin{deluxetable}{llll}
\label{tab:sources}
\tabletypesize{\scriptsize}
\tablewidth{0pt}
\tablecaption{Compact 2~mm Sources Near the Radio Arc}
\tablehead{
\colhead{Name} &
\colhead{$l$ [deg]} & 
\colhead{$b$ [deg]} &
\colhead{Other designation}
}
\startdata
A & 0.170 & -0.084 & N3 molecular cloud\\
B & 0.161 & -0.066 & Pistol Nebula\\
C & 0.209 & -0.001 & N2\\
D & 0.098 & -0.051 & N1\\
E & 0.078 & -0.093 & GPSR5 0.077-0.092\\
F & 0.097 & -0.165 & GPSR5 0.099-0.165 \& GPSR5 0.099-0.167\\
\enddata
\end{deluxetable}

\subsubsection{Source A}
Source A is detected at
160 -- 350 $\mu$m in the far-IR {\it Herschel} data, and in the 
450 and 850~$\mu$m SCUBA-2 results presented by \cite{Parsons:2018}, and the 
870~$\mu$m LABOCA data of \cite{Schuller:2009}, though flux measurements are 
highly uncertain because of the relatively low angular resolution of the data 
and the complexity of the background and surrounding sources.
The 2~mm flux density of source A, $0.13\pm0.02$ Jy, 
is a plausible extrapolation of the far-IR emission (Fig.~\ref{fig:point_spectra1}). 
Radio emission from N3 is seen at 8~mm -- 20~cm in Figure~\ref{fig:points} and 
the work of \cite{Ludovici:2016}. 
At 90~cm, the resolution is too low and N3 is too faint to be distinguished.
\cite{Ludovici:2016} show that the radio spectrum of N3 breaks to a steep 
$\alpha = -0.8$ spectral index in the $10-36$~GHz ($8-30$~mm) range. 
Thus the 2~mm flux density of Source A is more than an order of magnitude 
brighter than the extrapolated radio flux density of N3.
The radio source N3 appears to be associated with a small molecular 
cloud \citep{Tsuboi:2011,Ludovici:2016} that lies $\sim9''$ north (Galactic) of N3. 
Though the GISMO beam is larger than this separation, Source A is slightly 
better aligned with this cloud than N3. Cold dust in the molecular cloud would 
explain the far-IR emission, but for reasons unclear, the extrapolation of the far-IR 
emission is less effective at removing the 2~mm emission of this cloud than it is for 
the general ISM. Such a residual is unusual, but far from unique in the GC region.
Experimentation with various alternate selections of the background 
suggest that residual emission of the superimposed NTF cannot contribute 
more than 50\% of the flux density of source A.

There is no clear indication of source A at 1.25 -- 70~$\mu$m 
although there is a point source,
2MASS J17462113-2850023, which is offset by $1.4''$ from the location of N3, and 
has very red colors at $\lambda < 3$~$\mu$m, but is blue at $\lambda > 3$~$\mu$m,
which is consistent with typical foreground stars.  Because the N3 radio source 
remains point-like at $\sim0.2''$ resolution \citep{Ludovici:2016}, we believe it 
unlikely to be associated with this apparent foreground star.

\subsubsection{Source B}
The 2~mm source B is clearly associated with the brightest (eastern) parts of the Pistol Nebula. 
There is no shorter wavelength emission from this source until 70~$\mu$m, where emission is centered
closer to the Pistol Star itself. The entire Pistol Nebula is saturated at 24~$\mu$m, but is clearly resolved at 8~$\mu$m.
The overall SED is consistent with ionized gas (free-free radio emission) and very warm dust
(mid-IR emission) of material in the strong radiation fields of the Pistol Star and the Quintuplet Cluster
\citep{Moneti:1999,Figer:1999,Rodriguez-Fernandez:2001}.
The radiation and winds of these stars would be effective at clearing the region of any parental 
molecular cloud and associated far-IR emission \citep{Rodriguez-Fernandez:2001,Simpson:2007,Simpson:2018}.

\subsubsection{Source C}
At the location of source C, 3.6 - 24 $\mu$m emission shows a red point source surrounded by diffuse 
emission and an adjacent IR dark cloud (IRDC). The blended point and diffuse source emission is saturated at 
24 $\mu$m, but the IRDC is still seen in absorption. In progressing
from 70~$\mu$m to 1~mm the centroid of the emission shifts from the point source to the IRDC.  This far-IR emission 
can be fit by a modified blackbody with $T = 25.8$~K and $\beta = 1.98$ (Fig. \ref{fig:point_spectra1}), in 
agreement with results from \cite{Guzman:2015}. 
At 2~mm - 20~cm the centroid 
is back at the location of the point source. The point source was investigated and rejected as a possible YSO,
as the mid-IR spectrum did not show a 15~$\mu$m CO$_2$ ice feature with a 15.4~$\mu$m shoulder 
\citep{An:2011}. \cite{Yusef-Zadeh:1986} referred to the source as N2, and concluded it is a relatively compact
\ion{H}{2} region surrounding a massive star. A class II methanol maser is
located in this region \citep[e.g.][]{Caswell:1996,Yusef-Zadeh:2009}.

\subsubsection{Source D}
The 2~mm source D is associated with the radio source N1, referred to as an \ion{H}{2} region by \cite{Lazio:2008}.
\cite{Butterfield:2018} note that the source is thermal, based on a P$\alpha$ counterpart in the survey by \cite{Wang:2010}.
There is a relatively faint red point source here at 3.6 - 8~$\mu$m. It is saturated at 24~$\mu$m, but not seen again until
2~mm - 20~cm. The full resolution 6~cm radio image indicates an offset of several arcseconds between the radio
source and the mid-IR source. This is similar to the case of source C, but here there is no indication of associated diffuse emission. 

\subsubsection{Source E}
The 2~mm source E appears as a fainter version of source C, having both point-like and diffuse components,
and a generally similar SED.

\subsubsection{Source F}
Source F, though unresolved by GISMO, is actually 2 sources: a point source (GPSR5 0.099-0.165),
and a small region of more diffuse emission (GPSR5 0.099-0.167). 
These are also identified as 
sources 19 and 21 in \cite{Downes:1979}. A weak H$_2$O maser at $V_{\rm LSR}=0$ 
at the location of source 19 suggests that this source may be 
closer than the Galactic center region \citep{Guesten:1983}.
\cite{Becker:1994} associate
these sources with IRAS 17432--2855, which has IR colors characteristic of an ultracompact \ion{H}{2} region (UCHII).
The diffuse emission source appears as a small bubble to the southeast of the point source at 5.8 and 8~$\mu$m, 
cataloged as CN~3 by \cite{Churchwell:2007}. At 24~$\mu$m and 6~cm (full resolution) 
the diffuse emission is more concentrated at the center of the bubble. 
The far-IR -- mm emission is clear and seems associated with the point source, despite 
being blended at the angular resolution of the {\it Herschel} data. The emission can be  
fit by a modified blackbody with $T = 28.3$~K and $\beta = 2.03$ (Fig. \ref{fig:point_spectra1}). 
This source is notably bright at 1~mm. With increasing 
wavelength from 2~mm to 20~cm the centroid of the emission shifts from the location of the point source to the location 
of CN~3.
An ISO SWS spectrum of the Radio Arc Bubble was taken near this location \citep{Levine:1999,Rodriguez-Fernandez:2001}, 
but the ISO aperture does not include either of the sources, despite the stated association of this observation with 
GPSR5 0.099-0.167 by \cite{Giveon:2002}.

\subsubsection{General}

\cite{Pound:2018} report 3~mm detections of sources C, E, and F (resolved into 2 sources) 
as compact or unresolved objects (resolution = $7.62''\times 3.51''$) 
with 3~mm flux densities that are similar to or lower than 
our 2~mm measurements. In contrast they report sources A (N3) and B (the Pistol Nebula) as extended
sources and find flux densities that are $\sim2-5$ times higher than our 2~mm measurements.
Integration of the 2~mm GISMO map over the larger aperture used by \cite{Pound:2018}
does yield a substantially high flux density for source A (N3), though this includes some emission that is 
not clearly related to source A (e.g. part of the NTF) and more importantly requires a 
more distant and fainter region for background estimation which may not properly represent 
the background at N3.

The diffuse emission and far-IR properties suggest that sources C, E, and F are associated with star formation, and are likely
a blend of hot dust and free-free emission surrounding a compact \ion{H}{2} region and cold dust emission from 
surrounding molecular clouds. Source D would also be consistent with a compact \ion{H}{2} region (ionized gas and hot dust)
but shows no indication of emission from cold dust, suggesting a smaller (or no) remaining molecular cloud. 

\begin{figure}[t] 
   \centering
   \includegraphics[width=3.5in]{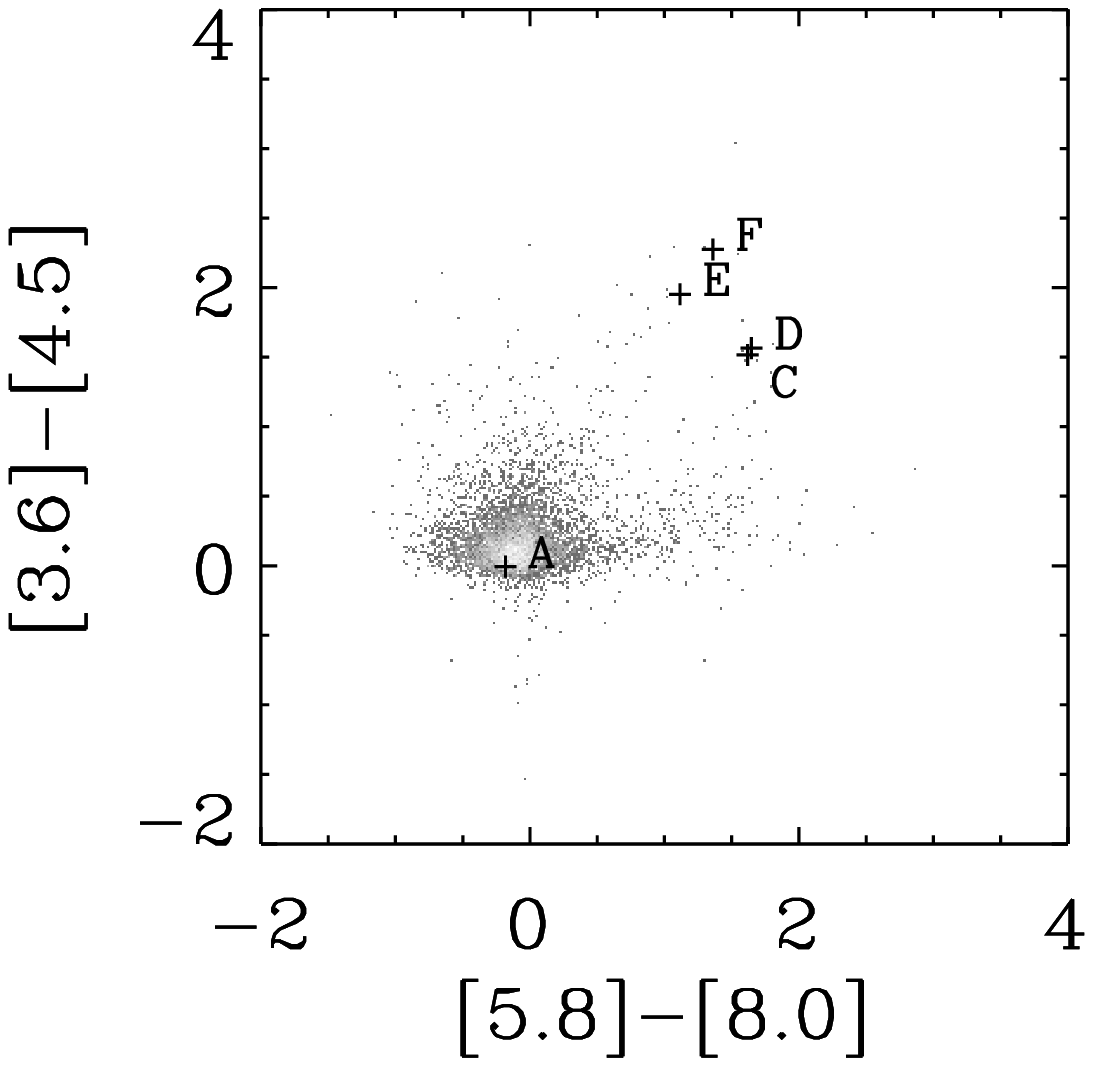}
   \includegraphics[width=3.5in]{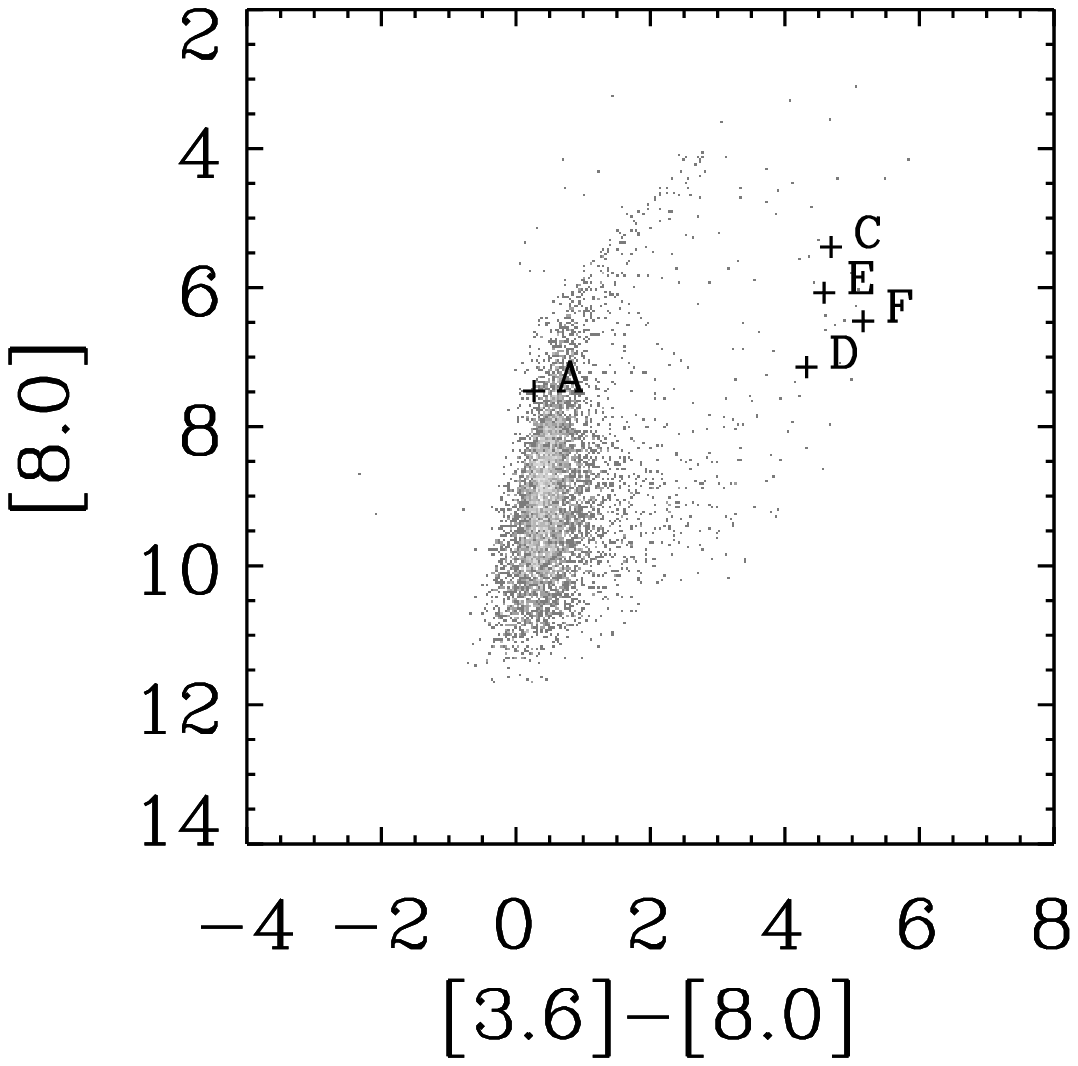}
   \caption{GLIMPSE II Archive color-color and color-magnitude plots
   for all sources within $10'$ of source A. These plots show that the 
   point labeled A is likely to be a foreground star with little reddening
   that coincidentally is along the line of sight to the 2~mm source A.
   The other sources have colors that are typical of star forming regions.
   \label{fig:point_colors}}
\end{figure}

The color-color and color-magnitude diagrams for all sources within $10'$ of source A
in the GLIMPSE II Archive are shown in Figure \ref{fig:point_colors}.
The locations of sources C and D in the IRAC color-color diagram
are consistent with those of the UCHII regions studied by 
\cite{de-La-Fuente:2009}. Sources E and F have somewhat redder $[3.6]-[4.5]$ colors
than the UCHII regions, but that could be an effect of stronger foreground 
reddening for lines of sight toward the Galactic Center. The more extended 
compact \ion{H}{2} regions studied by \cite{Phillips:2008} have bluer 
$[3.6]-[4.5]$ colors on average than the UCHII regions. The 
colors of sources C, D, E, and F are also consistent with those of 
candidate YSOs \citep[e.g.][]{Ramirez:2008}.
The colors of 2MASS J17462113-2850023 (marked as A) which 
lies within $1.5''$ of source A, are consistent with those of a typical red giant. 
This star's location in the color-magnitude diagram is clearly to the blue side of most of the 
cataloged stars in the field, suggesting that it may be a relatively nearby foreground 
star with low extinction.

%===================================
\section{Discussion} \label{sec:discussion}
%===================================
\subsection{Other Observations of the NTF}
A portion of the Radio Arc had previously been observed at 150~GHz (2~mm) by \cite{Reich:2000}.
In the region southward from the Sickle, they detected rather clumpy emission along the Radio Arc, 
but displaced slightly to the west. The brightest clump, at $b \sim -0.13\arcdeg$, was reported
to have a peak brightness near 300 mJy beam$^{-1}$ ($21.5''$), compared to a peak brightness 
in the Sickle region of $\sim170$ mJy beam$^{-1}$. The GISMO results present a very different picture.
We find that the bright NTF has a relatively smooth profile, which has a brightness of only 
$\sim 40$ mJy beam$^{-1}$ ($21''$) at $b \sim -0.13\arcdeg$, which is several times fainter than
the Sickle brightness of $\sim200$ mJy beam$^{-1}$.  Thus the GISMO results are 
roughly consistent with the spectral break to $\alpha = -1$ at $\nu > 40$ GHz 
as reported by \cite{Sofue:1999a}.
We cannot explain why the earlier observations
seem to show structure that is absent in the GISMO observations. The GISMO observations
do show some bright clumpy emission near the Radio Arc that is well matched as an extension 
of the far-IR spectrum of dust in molecular clouds, but these do not correspond to features in 
the \cite{Reich:2000} observations. The clear similarities
between the GISMO observations and other wider surveys at longer and shorter wavelengths 
provides confidence that the GISMO results are more reliable. 

The structure seen in the GISMO observations is also very well matched to 
the 3~mm observations by \cite{Pound:2018}. The 3~mm interferometric observations 
have higher angular resolution, but are at lower signal to noise and are insensitive to 
diffuse large scale emission. The 3~mm data are much less contaminated by thermal
emission from dust in molecular clouds, and therefore the resemblance to the 
2~mm data is clearer after the contributions of dust emission have been modeled
and subtracted \citep[see][]{Arendt:2019}.

\cite{Sofue:1992} report a nondetection of the NTF at 43~GHz (7~mm) using the
Nobeyama Millimeter-wave Array at $<1.4$~MJy~sr$^{-1}$ with a $16''\times10''$
synthesized beam. However they note that single-dish observations at the 
same wavelength (and $38''$ resolution) yield a mean brightness of 
16.3~MJy~sr$^{-1}$. This latter brightness is consistent with the 
other measurements shown in Figure \ref{fig:filament_sed}.

The GISMO observations of the NTF provide further support for its 
spectral turnover at $\sim 40$~GHz that was reported by \cite{Sofue:1999}.
The relative brightness of the central NTF and the rest of the Radio Arc, 
which is only barely visible on the western side of the NTF, indicates that 
the bulk of the Radio Arc has a steeper spectrum or lower-frequency turnover.
\cite{Reich:1988} explained the 843~MHz to 43.25~GHz morphology 
and spectral index of the Radio Arc as a broad component with $\alpha = -0.2$
and a narrower component with $\alpha = 0.3$. As these are determined over 
such a wide frequency range, they are not inconsistent with both components
rolling over to a much steeper spectrum at frequencies $\gtrsim 40$~GHz.

\subsection{Emission from the Radio Arc NTF}
Given that the 2~mm data confirm a distinct break in the radio spectrum,
it seems unlikely that free-free emission in the vicinity (or foreground)
of the filaments would contribute much to the 2~mm emission. 
Thus the 2~mm emission effectively sets an upper 
limit on the free-free at longer wavelengths
which is sufficiently low to alleviate the concern that free-free emission
may contribute to the flatness of the NTF radio spectrum \citep{Yusef-Zadeh:2003}. 
Dominance by nonthermal synchrotron emission at 2~mm is also supported by 
the finding that the Radio Arc filaments are one of the few polarized regions 
in low resolution 2 and 3~mm observations of the Galactic 
Center \citep{Culverhouse:2011}.
Furthermore, since our measurements are made using local background 
subtraction along the filament, any diffuse free-free emission should 
not contribute to measured brightness unless the emitting gas is also 
highly confined to the same flux tube.

The key features of the 2~mm emission from the NTF are its relatively 
steep spectral index compared to the 6 and 20~cm emission, and the 
observation that this spectral index steepens from north to south along the 
filament. To gain some insight to the meaning of these results, we 
compare the observations to theoretically-motivated synchrotron spectra.

Consider a source that injects high-energy electrons into a 
magnetically-confined filament with a cross sectional radius $R$. We assume that 
the energy spectrum of the injected electrons remains constant in time, so that 
the injection rate of electrons 
per unit volume 
can be represented by the product of two independent functions
\begin{equation}
\label{Qdot}
dQ(E_0, t_0) = k(t_0)\, n(E_0) dE_0
\end{equation}
where $n(E_0)dE_0$ is the number density of electrons with energies between $E_0$ and $E_0+dE_0$, and $k(t_0)$ is the injection rate at time $t_0$.

We assume that the energy spectrum of the electrons is given by a power law 
with an index $p$ at energies between $E_1$ and $E_2$, such that 
\begin{equation}
\label{}
n(E_0)dE_0 = \frac{\cal E}{\left<E\right>}\, \xi\, E_0^{-p}\, dE_0
\end{equation}
where $\left<E\right>=\xi\,  \int E_0^{-p+1}\, dE_0$ and ${\cal E}$ are, respectively, the average energy of the injected electrons, and the total electron energy density  over the $E_1-E_2$ energy range, 
and the coefficient $\xi$ is a normalization constant:   $\xi = \left[  \int E_0^{-p}\, dE_0\right]^{-1}$.

Following their injection, the electrons lose energy by synchrotron radiation at a rate given by \cite{Rybicki:1986}
\begin{eqnarray}
\label{rad}
-\frac{dE}{dt} & = & \frac{4}{3} \sigma_{\rm T}\, c\, U_B\, \beta^2\, \gamma^2 \nonumber \\
& \equiv & E / \tau_{\rm sync}
\end{eqnarray}
where $\sigma_{\rm T} = 6.65\times10^{-25}$~cm$^2$ is the Thomson cross section, $U_B=B^2/8\pi$ is the magnetic energy density, $B$ is the magnetic field strength, $c$ is the speed of light, and $\gamma=E/mc^2 = (1-\beta^2)^{-1/2}$, where $\beta = v/c \approx 1$,  $v$ is the electron velocity, and where the energy loss is averaged over pitch angles ($\left<\sin^2{\alpha}\right>=2/3$). The 
parameter $\tau_{\rm sync}$ defined in Equation (\ref{rad}) is the 
average synchrotron lifetime of the electron.

The specific emissivity of the synchrotron emission per unit volume at frequency $\nu$ and time $t$ is then given by:
\begin{eqnarray}
\label{sync}
j_{\nu}(\nu, t) & = & \int_0^t\, k(t_0) \int_{E_1}^{E_2}\ P_{\nu}[\nu, B, E(t-t_0)]\, n(E_0)\, dE_0\, dt_0 \nonumber \\
&  & \\
& = & \int_0^t\, \xi\, \frac{{\cal E}}{\left<E\right>}\ k(t_0) \int_{E_1}^{E_2}\ P_{\nu}[\nu, B, E(t-t_0)]\, E_0^{-p}\, dE_0\, dt_0 \nonumber
\end{eqnarray}
where the synchrotron power per unit frequency interval for a single electron is 
\begin{equation}
\label{pnu}
P_{\nu}(\nu, B, E) = \frac{2}{3}\, \sigma_{\rm T}\, c\, \frac{U_B}{\nu_{\rm c}}\, \gamma^2\ \frac{9\sqrt{3}}{4\pi}\, F(x) \ ,
\end{equation}
$\nu_{\rm c}=(3/4\pi)\, (eB/mc)\, \gamma^2\, \sin(\alpha)$ is a critical frequency, $e$ is the electron charge, $\alpha$ is the pitch angle between the electron velocity and the magnetic field, and $F(x)$ is a function of $x=\nu/\nu_c$ that can be approximated by (D. Kazanas, private communications)
\begin{equation}
\label{fx}
F(x) = 2.15\, x^{1/3}\, e^{-x}\ .
\end{equation}
This approximation is accurate within $\sim20\%$ for $x<100$.
The spectrum $P_{\nu}$ is calculated at the energy $E(t-t_0)$, the energy of the electron at time $t-t_0$ after its injection, which is given by
\begin{equation}
\label{en}
E(t-t_0) = \frac{E_0}{1+(t-t_0)/\tau_{\rm sync}} \ .
\end{equation}
The observed specific flux is given by an integral over the emitting volume 
 \begin{eqnarray}
\label{fnu}
F_{\nu}(\nu, t) & = & \frac{1}{4 \pi D^2}\ \int j_{\nu}(\nu, t)\ dV \nonumber \\
 & = & \frac{\Omega}{4 \pi}\,  j_{\nu}(\nu, t)\, \ell
\end{eqnarray} 
where $\Omega$ is the beam size, and $\ell = 4R/3$ is the effective path length through the filament.

To fit the observed spectrum we adopted the following model parameters: $E_1$
and $E_2$ equal to $10^6$ and $5\times10^{10}$~eV, respectively, and 
assume the magnetic field is uniform
with a strength of $B = 10^{-3}$~G \citep[e.g.][]{Yusef-Zadeh:1987} 
along the filament. The flatness
of the synchrotron spectrum at $\gtrsim 6$~cm suggests a flat injection
spectrum of the electrons with a power-law index of $p=0.92$. With this choice
of parameters, the average electron energy is $\left<E\right> = 5.2\times 10^9$~eV,  with a
corresponding synchrotron lifetime of 3200~y. We took the radius of the filament
to be $R = 0.15$~pc \citep{Morris:1985}. Our examination is limited to the range
$-0.16\arcdeg < b < -0.05\arcdeg$ where the filament is best seen,
corresponding to a length of $4.8\times 10^{19}$~cm, 
assuming a distance of 8.18~kpc \citep{Abuter:2019}.

\begin{figure}[t] 
   \centering
   \includegraphics[width=3.5in]{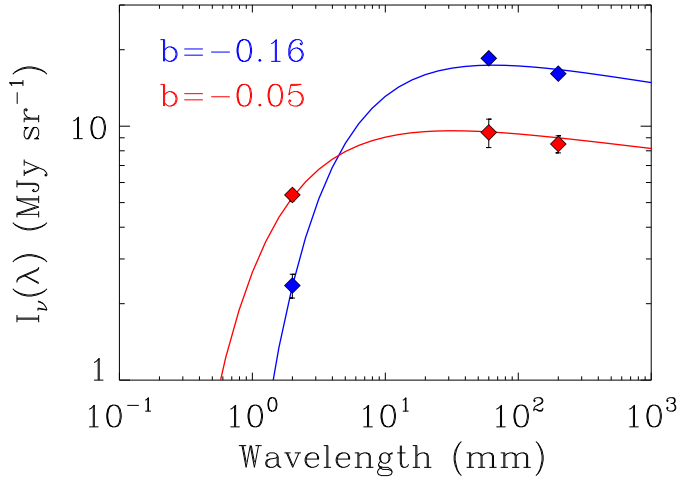} 
   \caption{Fits of a simple synchrotron model to the spectra
   of the NTF at contrasting locations. The location at $b=-0.05\arcdeg$ 
   is near the Sickle. The location at $b=0.16\arcdeg$ is
   toward the south end of the filament, where the 2~mm to 6 and 20~cm 
   spectral indices are distinctly steeper. See text for details.
   \label{fig:modelfits}}
\end{figure}

We assume that the electrons are injected into the
filament at some time $t=t_0$, and that their energies evolve 
due to synchrotron losses as described in Equation (7). 
Fitting the data allows us to estimate the time elapsed since their injection. Figure~\ref{fig:modelfits}
depicts the fits of this model at the opposite endpoints of this 
segment of the filament, where the spectral indices are distinctly different.
The fits suggest ages of 2500 and  5000~y and electron 
energy densities of 1.3 and $2.1\times 10^{-10}$~erg~cm$^{-3}$ at the 
north and south ends respectively.
  
If the electrons are injected into the filament near the 
northern position, and the age difference corresponds to the 
electron diffusion time to the southern position, then the implied
velocity is $\sim 6100$~km~s$^{-1}$. As in an ionized medium 
this velocity is limited to the Alfv\'en velocity, 
$v_{\rm A} \approx 2\times10^{11} B\, n_{\rm H}^{-1/2}$~cm~s$^{-1}$
\citep{Wentzel:1974,Longair:1981},
we derive an average density of nuclei in the filaments of
$n_{\rm H} = 0.1$~cm$^{-3}$.

The total energy contained in this segment of the filament is 
$E_{\rm tot} \approx 5.7\times10^{45}$~erg. If the filament is not 
a transient phenomenon, then the luminosity
required to sustain the emission from the filament is approximately given by
$E_{\rm tot}/\tau_{\rm sync} \approx 15$~L$_{\sun}$.  This luminosity is very similar to the independently obtained value obtained by \cite{Serabyn:1994}. 
However, it depends on our choice of the magnetic field 
strength and the value of $E_2$. For example, for a fixed magnetic 
field strength, we find that the filament's luminosity scales 
as $\sim E_2^{1.5}$. A detailed examination of the model 
scenario and parameter space will be a subject of future studies.

A number of models for electron acceleration giving rise to the synchrotron emission in the 
Galactic center filament were summarized by \cite{Morris:1996a}.
One hypothesis by \cite{Serabyn:1994}
invoked acceleration by magnetic field line reconnection at the surface of the molecular 
cloud that is ionized by UV radiation from the Quintuplet cluster.  
They had found reconnection plausible because the magnetic fields inside and outside 
of the molecular cloud are orthogonal to each other, and they can be mixed 
and reconnecting in the turbulent, ionized cloud surface. They also noted that 
there are particular molecular clumps associated with each of the primary bundles 
of filaments in the Radio Arc (see Figure 2 of that paper). That, and the 
fact that several of the filaments undergo a discontinuity in either brightness or 
direction (slightly in the case of direction) at the location of the ionization 
front, suggested an association of the production sites for relativistic particles 
with the ionized surfaces of the molecular clumps.
However, we do note that at 90~cm, the Sickle appears in 
absorption against the Radio Arc filaments, suggesting that it lies in front,
at least in part (Figure \ref{fig:radio}). This was noted by 
\cite{Anantharamaiah:1991}, who also point out likely absorption by the Pistol
Nebula as well (Source B in Figure \ref{fig:radio}).
There are other clear cases 
where an ionized surface on a molecular cloud is associated with a filament,
as studied by \citep{Uchida:1996} and \cite{Staguhn:1998}.

Alternately, production of the relativistic electrons may arise in shocked stellar winds as 
proposed by \cite{Rosner:1996} and \cite{Yusef-Zadeh:2003}. 
In the case of single adiabatic diffusive shock acceleration, the spectral index 
is $\alpha = -0.5$. Having multiple shocks flattens the distribution of the 
energy spectrum of the particles. Winds from different stellar sources may be 
responsible if they collide and get shocked multiple times \citep{Pope:1994}.
The shocks would be due to the collective winds from all the stars in the Quintuplet Cluster, 
which is more compact than the scale of the ionization front/shock (i.e., the Sickle). 
While the Radio Arc NTFs have a flat spectrum at longer wavelengths, most 
other nonthermal filaments have steep spectra \citep{Law:2008a}. 
However, this model does not give an obvious reason why the prominent filament bundles 
in the Radio Arc are associated with molecular clumps. From this, one might expect that 
electrons would be accelerated across the shock all the way along the Sickle, rather than 
in just a few choice places.

The interaction of a Galactic wind with dense clouds has also been 
proposed by \cite{Shore:1999} as a means of creating NTFs, with more detailed
development by \citep{Dahlburg:2002}. 
Given the large number of filaments in the Galactic center, 
a global wind is needed and at the sites where the wind gets shocked, producing 
relativistic particles. In the case of the Radio Arc, the wind could 
come from the Quintuplet cluster.

%===================================
\section{Summary} \label{sec:summary}
%===================================

The brightest NTF in the Radio Arc is detected at 2~mm with a 
brightness that is consistent with a 
steep spectral index ($\alpha ~\approx -1.5$) at 2~mm.
The steepening 2~mm spectral index as a function of location along 
the NTF points to the Sickle (rather than point source N3) being 
directly related to the origin on the NTF electrons.
Interpretation of the changing spectral index as 
the aging of populations of relativistic electrons implies
timescales of $\sim5000$~y and velocities of $\sim6100$~km~s$^{-1}$ 
for diffusion of relativistic electrons along the NTF.
The rest of the Radio Arc is only marginally detected.

We detect a 2~mm point source (at $21''$ resolution) 
very near the location of the N3 radio point source, but 
examination of the data across a wide range of wavelengths indicates 
the 2~mm emission is likely dust emission from an adjacent molecular cloud.
We determine that a near-IR ($1-8$ $\mu$m) point source
coincident with this 2~mm point source
is actually a foreground star along the same line of sight, 
and is unrelated to the radio source N3 as well.
Other nearby compact 2~mm sources, unlike the source near N3, 
show extended emission in higher resolution mid-IR (3.6 - 24~$\mu$m) 
data and are mostly associated with compact \ion{H}{2} regions and other 
markers of star formation. 

This report covers only a portion of the 
Galactic Center region mapped by GISMO. Other regions of interest are 
the dust properties of the ISM and dense molecular clouds in the GC region 
\citep{Arendt:2019}, thermal and non-thermal emission in the vicinity of Sgr A
and Sgr A* itself, thermal emission regions near Sgr C and Sgr B1, and 
the dense molecular clouds and embedded UCHII regions in Sgr B2.

\acknowledgments
We would like to thank 
Carsten Kramer, Santiago Navarro, David John, Albrecht Sievers,
and the entire IRAM Granada staff for their support during the instrument 
installation and observations. 
We thank the referee for comments that improved the clarity and 
utility of the manuscript.
IRAM is supported by INSU/CNRS (France), 
MPG (Germany), and IGN (Spain). This work was supported 
through NSF ATI grants 1020981 and 1106284.

\facilities{{\it Spitzer} (IRAC and MIPS), {\it Herschel} (PACS and SPIRE), IRAM:30~m (GISMO), VLA}

\software{CRUSH \citep{Kovacs:2008}, IDLASTRO \citep{Landsman:1995}, SAO ds9 \citep{Joye:2003}}\\

{\it \large{Note added in proof:}}~~
\cite{Heywood:2019} have used the MeerKAT radio telescope to produce an
extremely wide and sensitive 1.284 GHz map of the Galactic Center region. The
map reveals greater detail and extent of the radio lobe \citep[e.g.][]{Law:2010}, both N
and S of the Galactic plane. Many of the NTFs appear to be coherent with the
walls of this radio bubble. \cite{Heywood:2019} suggest NTFs outside 
the radio bubble may result from cosmic ray advancing ahead of a shock front.
Thus there may even be a common origin for all the NTFs, rather
than a locally-driven process that is repeated independently throughout the CMZ.
%~~
%
%~~
%
%~~
%
%~~

\bibliography{gismo_gc}

\end{document}